\newcommand\apjcls{1}
\newcommand\aastexcls{2}
\newcommand\othercls{3}

\newcommand\papercls{\aastexcls}
\documentclass[tighten, times, twocolumn]{aastex62}  % ApJ look-alike
\if\papercls \apjcls
\usepackage{apjfonts}
\else\if\papercls \othercls
\usepackage{epsfig}
\usepackage{margin}
\usepackage{times}
\fi\fi
\usepackage{ifthen}
\usepackage{natbib}
\usepackage{amssymb, amsmath}
\usepackage{appendix}
\usepackage{etoolbox}
\usepackage[T1]{fontenc}
\usepackage{paralist}

\if\papercls \apjcls
\newcommand\aas{\ref@jnl{AAS Meeting Abstracts}}% *** added by jh
\newcommand\dps{\ref@jnl{AAS/DPS Meeting Abstracts}}% *** added by jh
\newcommand\maps{\ref@jnl{MAPS}}% *** added by jh
\else\if\papercls \othercls
\usepackage{astjnlabbrev-jh}
\fi\fi

\if\papercls \aastexcls
\hypersetup{citecolor=blue, % color for \cite{...} links
            linkcolor=blue, % color for \ref{...} links
            menucolor=blue, % color for Acrobat menu buttons
            urlcolor=blue}  % color for \url{...} links
\else
\usepackage[%pdftex,      % hyper-references for pdflatex
bookmarks=true,           %%% generate bookmarks ...
bookmarksnumbered=true,   %%% ... with numbers
colorlinks=true,          % links are colored
citecolor=blue,           % color for \cite{...} links
linkcolor=blue,           % color for \ref{...} links
menucolor=blue,           % color for Acrobat menu buttons
urlcolor=blue,            % color for \url{...} links
linkbordercolor={0 0 1},  %%% blue frames around links
pdfborder={0 0 1},
frenchlinks=true]{hyperref}
\fi

\if\papercls \othercls

\else

\fi

\providecommand{\adsurl}[1]{\href{#1}{ADS}}

\makeatletter
\patchcmd{\NAT@citex}
  {\@citea\NAT@hyper@{%
     \NAT@nmfmt{\NAT@nm}%
     \hyper@natlinkbreak{\NAT@aysep\NAT@spacechar}{\@citeb\@extra@b@citeb}%
     \NAT@date}}
  {\@citea\NAT@nmfmt{\NAT@nm}%
   \NAT@aysep\NAT@spacechar\NAT@hyper@{\NAT@date}}{}{}

\patchcmd{\NAT@citex}
  {\@citea\NAT@hyper@{%
     \NAT@nmfmt{\NAT@nm}%
     \hyper@natlinkbreak{\NAT@spacechar\NAT@@open\if*#1*\else#1\NAT@spacechar\fi}%
       {\@citeb\@extra@b@citeb}%
     \NAT@date}}
  {\@citea\NAT@nmfmt{\NAT@nm}%
   \NAT@spacechar\NAT@@open\if*#1*\else#1\NAT@spacechar\fi\NAT@hyper@{\NAT@date}}
  {}{}
\makeatother

\makeatletter
\DeclareRobustCommand{\lowcase}[1]{\@lowcase#1\@nil}
\def\@lowcase#1\@nil{\if\relax#1\relax\else\MakeLowercase{#1}\fi}
\makeatother

\DeclareSymbolFont{UPM}{U}{eur}{m}{n}
\DeclareMathSymbol{\umu}{0}{UPM}{"16}
\let\oldumu=\umu
\renewcommand\umu{\ifmmode\oldumu\else\math{\oldumu}\fi}

\if\papercls \othercls

\else

\fi

\let\oldsim=\sim
\renewcommand\sim{\ifmmode\oldsim\else\math{\oldsim}\fi}
\let\oldpm=\pm
\renewcommand\pm{\ifmmode\oldpm\else\math{\oldpm}\fi}
\newcommand\by{\ifmmode\times\else\math{\times}\fi}
\newbox{\wdbox}
\renewcommand\c{\setbox\wdbox=\hbox{,}\hspace{\wd\wdbox}}
\renewcommand\i{\setbox\wdbox=\hbox{i}\hspace{\wd\wdbox}}

\newcount\timect
\newcount\hourct
\newcount\minct
\newcommand\now{\timect=\time \divide\timect by 60
         \hourct=\timect \multiply\hourct by 60
         \minct=\time \advance\minct by -\hourct
         \number\timect:\ifnum \minct < 10 0\fi\number\minct}

\catcode`@=11

\newcommand\comment[1]{}

\newcommand\commenton{\catcode`\%=14}

\renewcommand\math[1]{$#1$}
\newcommand\mathshifton{\catcode`\$=3}

\let\atab=&
\newcommand\atabon{\catcode`\&=4}

\let\oldmsp=\sp
\let\oldmsb=\sb
\def\sp#1{\ifmmode
           \oldmsp{#1}%
         \else\strut\raise.85ex\hbox{\scriptsize #1}\fi}
\def\sb#1{\ifmmode
           \oldmsb{#1}%
         \else\strut\raise-.54ex\hbox{\scriptsize #1}\fi}
\newbox\@sp
\newbox\@sb
\def\sbp#1#2{\ifmmode%
           \oldmsb{#1}\oldmsp{#2}%
         \else
           \setbox\@sb=\hbox{\sb{#1}}%
           \setbox\@sp=\hbox{\sp{#2}}%
           \rlap{\copy\@sb}\copy\@sp
           \ifdim \wd\@sb >\wd\@sp
             \hskip -\wd\@sp \hskip \wd\@sb
           \fi
        \fi}
\def\msp#1{\ifmmode
           \oldmsp{#1}
         \else \math{\oldmsp{#1}}\fi}
\def\msb#1{\ifmmode
           \oldmsb{#1}
         \else \math{\oldmsb{#1}}\fi}

\def\supon{\catcode`\^=7}

\def\subon{\catcode`\_=8}

\def\supsubon{\supon \subon}

\newcommand\actcharon{\catcode`\~=13}

\newcommand\paramon{\catcode`\#=6}

\comment{And now to turn us totally on and off...}

\newcommand\reservedcharson{ \commenton  \mathshifton  \atabon  \supsubon 
                             \actcharon  \paramon}

\catcode`@=12
\reservedcharson

\if\papercls \apjcls

\else

\fi

\newcommand\chisq{\ifmmode{\chi\sp{2}}\else\math{\chi\sp{2}}\fi}
\newcommand\redchisq{\ifmmode{ \chi\sp{2}\sb{\rm red}}
                    \else\math{\chi\sp{2}\sb{\rm red}}\fi}
\newcommand\Teq{\ifmmode{T\sb{\rm eq}}\else$T$\sb{eq}\fi}
\newcommand\mjup{\ifmmode{M\sb{\rm Jup}}\else$M$\sb{Jup}\fi}
\newcommand\rjup{\ifmmode{R\sb{\rm Jup}}\else$R$\sb{Jup}\fi}
\newcommand\msun{\ifmmode{M\sb{\odot}}\else$M\sb{\odot}$\fi}
\newcommand\rsun{\ifmmode{R\sb{\odot}}\else$R\sb{\odot}$\fi}
\newcommand\mearth{\ifmmode{M\sb{\oplus}}\else$M\sb{\oplus}$\fi}
\newcommand\rearth{\ifmmode{R\sb{\oplus}}\else$R\sb{\oplus}$\fi}
\shorttitle{Hot Hydrogen Climates}
\shortauthors{Koll \& Cronin}

\begin{document}

\title{Hot Hydrogen Climates near the inner edge of the Habitable Zone}

%% AUTHOR/INSTITUTIONS FOR AASTEX6.1:
\author{Daniel~D.B.~Koll}
\affiliation{Department of Earth, Atmospheric and Planetary Sciences,
  Massachusetts Institute of Technology, Cambridge, MA, 20139.}

\author{Timothy~W.~Cronin}
\affiliation{Department of Earth, Atmospheric and Planetary Sciences,
  Massachusetts Institute of Technology, Cambridge, MA, 20139.}

\email{dkoll@mit.edu}

% %% Extra info for aastex:
% \received{Yesterday}
% \revised{Today}
% \accepted{Tonight}
% \published{Tomorrow}
% \submitjournal{AASJournal}

% \turnoffedit           % from AASTEX: turn off \edit markup?

\begin{abstract}
Young terrestrial planets can capture or outgas hydrogen-rich
atmospheres with tens to hundreds of bars of H$_2$, which persist
for 100 Myrs or longer.
Although the earliest habitable conditions on Earth and terrestrial
exoplanets could thus arise while the atmosphere is still dominated
by H$_2$, the climatic effects of H$_2$ remain poorly understood. 
Previous work showed that H$_2$ induces strong greenhouse warming
at the outer edge of the habitable zone.
Here we use a 1D radiative-convective model to show that H$_2$ also
leads to strong warming near the inner edge of the habitable zone.
Unlike H$_2$'s greenhouse warming at the outer edge, however, its
effect near the inner edge is driven by thermodynamics: H$_2$'s large
thermal scale height allows the atmosphere to store more water vapor than either a
pure-H$_2$O atmosphere or an atmosphere with a heavy background gas,
such as N$_2$ or CO$_2$, thereby amplifying
the greenhouse effect of H$_2$O.
Using idealized grey calculations, we then present a general
  argument for how different background gases affect the inner edge of
  the habitable zone. H$_2$ stands out for its ability to induce novel
  `Souffl\'e' climates, which further support its warming effect.
Our results show that if the earliest conditions on a planet
  near the inner edge of the habitable zone were H$_2$-rich, they were
  likely also hot: 1 bar of H$_2$ is sufficient to raise surface
  temperatures above 340 K, and 50 bar of H$_2$ are sufficient to
  raise surface temperatures above 450 K.
\end{abstract}

%http://journals.aas.org/authors/keywords2013.html

%%%%%%%%%%%%%%%%%%%%%%%%%%%%
\section{Introduction}
\label{introduction}

Hydrogen plays an important role in the early stages of terrestrial planet evolution.
Planets that form inside a nebular disk are likely to gravitationally capture a primordial H$_2$ atmosphere \citep{rafikov2006a,pierrehumbert2011a}.
Even after the disk disperses, planets can outgas H$_2$ during
the solidification of the planet's magma ocean
\citep{elkins-tanton2008,schaefer2010}, or can form H$_2$ through the
oxidation of iron-rich impactors during the giant impact
phase \citep{genda2017}.
The amount of H$_2$ generated varies with formation mechanism
but can be substantial, ranging from several to more than 1000 bars
\citep{pierrehumbert2011a, elkins-tanton2008,genda2017}.

H$_2$ eventually dissipates because its low molecular weight allows
it to escape to space. In the meantime, however, it can shape
a planet's earliest surface conditions.
To order of magnitude, we estimate that it takes $\sim$100 Myr for a
young Earth-like planet around a Sun-like star to lose 100
bars of H$_2$ to space (see Appendix A).
This timescale becomes longer if H$_2$ is produced or outgassed
  later, due to the rapid decrease over time in the XUV output of young Sun-like stars.
Our estimate agrees with the results of \citet{genda2017}, who
argued that early Earth could have had a $\sim$70 bar H$_2$
  atmosphere which gradually escaped over 200 Myr, produced by the
  oxidation of fragments from a giant impact.

These timescales are longer than the $1-10$ Myr needed to solidify a
magma ocean and condense a steam atmosphere
\citep{hamano2013a,lebrun2013}. They are also long enough
  to shape prebiotic chemistry and, potentially, for an early
  biosphere to develop \citep{lazcano1994,bell2015,betts2018}.
Some terrestrial planets, including ancient Earth, could
therefore undergo an early habitable phase with liquid water oceans
underneath a H$_2$ atmosphere.
What would such a climate be like?

The habitable zone has conventionally only been studied for an atmosphere
dominated by high mean-molecular-weight (MMW) background gases, in
particular N$_2$ and CO$_2$ \citep[e.g.][]{kasting1993,kopparapu2013b}.
There are relatively few studies that have considered the habitability
of H$_2$-rich planets, and these studies either focused on thick H$_2$ atmospheres
at the outer edge of the habitable zone or atmospheres with relatively
small amounts of H$_2$.
In both cases, hydrogen can have a significant effect on climate.
Thick H$_2$ envelopes, with tens of bars of H$_2$, generate a strong
greenhouse effect via H$_2$-H$_2$ collision-induced absorption that
can allow planets to remain habitable far beyond the outer edge
of the conventional N$_2$-CO$_2$ habitable zone
\citep{stevenson1999,pierrehumbert2011a,wordsworth2012}.
Similarly, even small amounts of H$_2$, on the order of 1-30\% molar
fraction, can induce greenhouse warming via H$_2$-N$_2$ or H$_2$-CO$_2$
collision-induced absorption.
H$_2$ could thus also help resolve the Faint Young Sun problem on
early Earth and early Mars, and could expand the outer edge of the habitable zone \citep{wordsworth2013b,ramirez2014c,batalha2015,ramirez2017e}.

Less attention has been paid to how a transient H$_2$ atmosphere,
  with 1 or more bars of H$_2$, would affect a planet's climate closer to the inner edge of the
habitable zone, which we consider in this work.
Unlike at the outer edge, where H$_2$'s radiative effect dominates, the
interaction between water vapor and the H$_2$ background becomes increasingly
important near the inner edge, which we define as roughly the region
where the incident stellar flux, or stellar constant, exceeds 1000 W m$^{-2}$
($\sim$70\% of the flux received by Earth today).
We explore these interactions with a 1D radiative-convective model
that represents a planet orbiting a Sun-like star.
To relate our results to conventional N$_2$-CO$_2$ habitable zone calculations, we
consider both high MMW and H$_2$-rich atmospheres (the model is
described in Section \ref{sec:methods}).
Our results fall into two categories. First, we show that H$_2$ has a
strong warming effect near the inner edge. This effect is
independent of H$_2$'s greenhouse effect and instead arises from
hydrogen's thermodynamic properties (Section \ref{sec:spectral}).
Second, we present a general consideration for how a planet's
background gases affect its climate, with a particular focus on the
approach to the runaway greenhouse. We show that a habitable planet
has distinct climate feedbacks with H$_2$ versus high-MMW backgrounds,
and that these feedbacks give rise to novel climate states in
H$_2$-rich atmospheres which further support H$_2$'s warming effect (Section \ref{sec:grey}).
Finally, we discuss our findings and its implications for the origin
of life in the early Solar System and around other stars (Section \ref{sec:discussion}).

%%%%%%%%%%%%%%%%%%%%%%%%%%%%
\section{Methods}
\label{sec:methods}

We the Python line-by-line RADiation model for planetary
atmosphereS (PyRADS), which is a 1D
model representing an atmosphere in radiative-convective
equilibrium.
We have validated PyRADS against other line-by-line calculations
  in the longwave \citep{koll2018d}, as well as the shortwave
  (Appendix B). The model code is open source and freely available on github\footnote{\href{https://github.com/ddbkoll/PyRADS}{github.com/ddbkoll/PyRADS-shortwave}}.

The atmosphere is composed of a dry background gas (e.g.,
N$_2$ or H$_2$) plus condensing water vapor. The amount of
atmospheric water vapor increases rapidly with temperature
following the Clausius-Clapeyron relation, so all atmospheres become
steam-dominated at high temperatures.  For example, the saturation
vapor pressure of water is $\sim 6 \times 10^{-3}$ bar  at the triple
point (273 K), whereas it is one bar at the boiling point (373 K).
A planet with 1 bar of background N$_2$ at a surface temperature of 273 K
therefore has an atmosphere that is largely made of N$_2$, whereas at
373 K the total surface pressure rises to 2 bar and the air near the ground
contains as many H$_2$O molecules as N$_2$ molecules.

To compute convective temperature profiles we use the general moist
adiabat, which is valid for both a dry atmosphere and a
steam-dominated atmosphere \citep{ding2016}.
All thermodynamic constants are taken from \cite{pierrehumbert2010}.
We choose a variable vertical resolution to ensure that we resolve the
upper atmosphere in the runaway limit, yet also have good resolution
near the surface at intermediate surface temperatures.
For the radiative transfer we perform two separate calculations in the longwave and
shortwave. 
Our longwave calculations cover the spectrum up to 5,000 
cm$^{-1}$ (2 $\mu$m) with a spectral resolution of 0.01 cm$^{-1}$.
In the shortwave we cover the spectrum between 1,000 and 110,000
cm$^{-1}$ (0.09-10$\mu$m) with a spectral resolution of 0.01 cm$^{-1}$
up to 30,000 cm$^{-1}$ and a resolution of 10 cm$^{-1}$ at higher
wavenumbers (see below).
We assume a Sun-like host star, with a solar spectrum taken from the VPL database
\citep{segura2003}, and a surface albedo of 0.12. To solve the
shortwave radiative transfer we use DISORT \citep{stamnes1988},
implemented via
PyDISORT\footnote{\href{https://github.com/chanGimeno/pyDISORT}{github.com/chanGimeno/pyDISORT}}. We
run DISORT at line-by-line resolution using 4 angular
streams. Sensitivity tests indicate that our results are essentially
identical if we use more angular streams.
To compute opacities we use a combination of H$_2$O lines from
HITRAN2016 \citep{gordon2017} and the more comprehensive ExoMol BT2 line
list \citep{barber2006}.
At the temperatures we are considering, the choice of line
list only has a minor impact on longwave fluxes because HITRAN2016 already
contains most strong absorption lines in the infrared. In contrast, shortwave fluxes
are sensitive to the choice of line list because HITRAN2016 does not
cover the UV even though the near and middle UV contains a moderate fraction of
the incoming stellar flux ($\sim 8\%$ for a Sun-like star).
As a compromise between computational tractability and complete
spectral coverage we adopt HITRAN2016 lines below 15,000 cm$^{-1}$ and
BT2 lines between 15,000 cm$^{-1}$ and 30,000 cm$^{-1}$.
Similarly, for computational tractability in our CO$_2$ calculations
we use the HITRAN2016 line list.
We use a Lorenz line shape throughout instead of a Voigt line
shape. This approximation means we do not resolve line cores, but the
resulting impact on OLR is small \citep{koll2018d}.

H$_2$O line broadening could behave differently in a N$_2$ versus a
H$_2$ atmosphere.  To explore this possibility we compared line widths
for H$_2$O-air broadening \citep{voronin2010} with line widths
for H$_2$O-H$_2$ broadening \citep{barton2017}. Despite our
expectation that collisional line width should scale with
intermolecular velocity \citep{pierrehumbert2010}, and hence with the
inverse square root of the broadening gas's molecular mass, H$_2$O line widths
are roughly the same for H$_2$ and N$_2$ broadening. In both cases the
widths of the strongest lines are about 0.1 cm$^{-1}$ atm$^{-1}$.
Although this invariance might be particular to H$_2$ and N$_2$, because line widths for H$_2$O-He
collisions are noticeably smaller \citep{barton2017}, we use
H$_2$O-air broadening parameters for both N$_2$ and H$_2$.

In addition to line opacities we include several other opacity
sources. We compute H$_2$O-H$_2$O and H$_2$O-foreign collision-induced
absorption (CIA) using the semi-empirical MTCKD continuum model
\citep{mlawer2012}. For H$_2$-H$_2$ CIA we use data from HITRAN. We
are not aware of any H$_2$O-H$_2$ CIA data, so we use H$_2$O-N$_2$ CIA
in both our N$_2$ and our H$_2$ calculations.
Our use of H$_2$O-N$_2$ data to mimic H$_2$O-H$_2$ CIA could induce
  errors but is reasonable given the lack of laboratory measurements: N$_2$
  and H$_2$ are both homonuclear diatomics, so to first order one might
  expect them to similarly interact with the H$_2$O molecule and its
  electric field moments.
For CO$_2$-CO$_2$ and CO$_2$-foreign CIA we use data from
\cite{pierrehumbert2010}.
For Rayleigh scattering
we use the N$_2$ and H$_2$O scattering cross-sections from
\cite{goldblatt2013}, the H$_2$ scattering cross-sections from
\cite{dalgarno1962}, and the CO$_2$ scattering cross-sections from
\cite{pierrehumbert2010}. We parameterize absorption in the UV by setting
the absorption cross-section of H$_2$O at frequencies higher
than 50,000 cm$^{-1}$ (0.2 $\mu$m) equal to a representative value
based on the MPI-Mainz Atlas, $\kappa_{UV}\approx 10^{-18}$ cm$^2$
molec$^{-1}$ $= 3\times 10^3$ m$^2$ kg$^{-1}$.
Consistent with our use of a 1D model we do not include effects that
are determined by 3D dynamics, such as clouds or changes in relative
humidity. Instead we assume clear-sky profiles and a relative
humidity of unity.
We also assume an all-adiabat atmosphere in which
convection reaches to the top of the atmosphere, and ignore the
stratosphere.
We discuss the impact of these modeling assumptions in Section \ref{sec:discussion}.
For our analytical results in Section \ref{sec:grey} we note that the definition of
optical thickness often includes a implicit average cosine
of the zenith angle, $\overline{\cos \theta}$, which accounts for the
angular distribution of longwave radiation \citep{pierrehumbert2010}. Our numerical calculations use $\overline{\cos \theta} =3/5$. To
reproduce our analytical values, optical thickness $\tau$ along
a vertical path should be divided by $\overline{\cos \theta}$.

Below we will present our results in terms of the planet's surface
temperature $T_s$ as a function of incoming stellar flux (Figure
\ref{fig:spectral}). To present our results in this way we first run
our model across a large grid of surface temperatures. The model
outputs albedo $\alpha$ and OLR, which then allows us to infer the
stellar flux $L_*$ that would be required to maintain a given surface temperature via
planetary energy balance, $L_* = 4 \mathrm{OLR}(T_s)/(1-\alpha(T_s))$.  We
refer to a climate state as stable if warming increases the planet's
OLR more than it increases the planet's absorbed shortwave flux:
\begin{eqnarray}
  \frac{1}{\mathrm{OLR}}\frac{d\mathrm{OLR}}{dT_s} +
  \frac{1}{(1-\alpha)}\frac{d\alpha}{dT_s} > 0.
\end{eqnarray}

%%%%%%%%%%%%%%%%%%%%%%%%%%%%
\section{H$_2$ warming near the inner edge.}
\label{sec:spectral}

\subsection{Spectral calculations}

\begin{figure*}[t]
\includegraphics[width=\linewidth, clip]{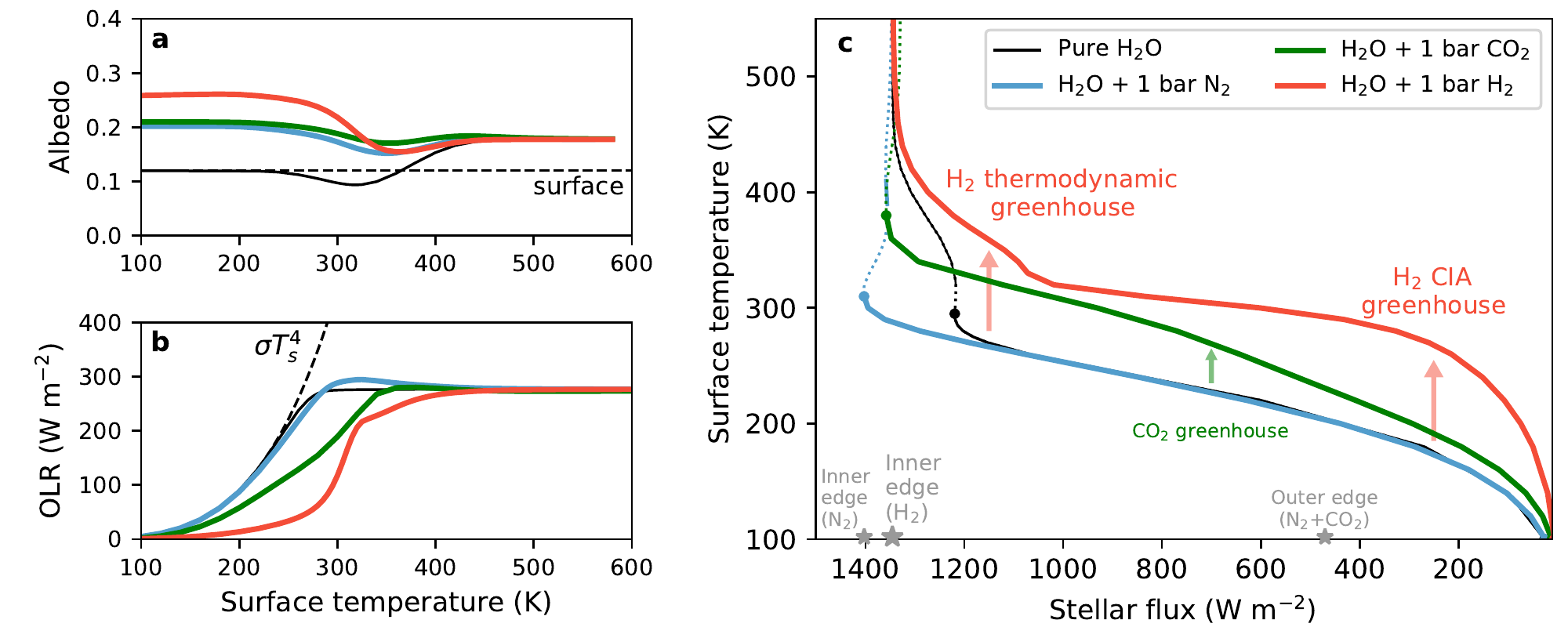}
\caption{The greenhouse effect of H$_2$:
  H$_2$ causes strong greenhouse warming near the outer
  edge of the habitable zone due to collision-induced absorption
  (CIA), but it also causes strong
  warming near the inner edge of the habitable zone due to its
  thermodynamic effect. In comparison, CO$_2$ is a relatively weaker
  greenhouse gas.
  (a,b) Albedo and outgoing longwave radiation (OLR) as a
  function of surface temperature. (c) Surface temperature as a
  function of received stellar flux. Solid
  lines are stable, dotted lines are unstable climate states.
  Grey stars show the conventional inner and outer edge of
    the habitable zone \citep{kopparapu2013b}, as well as the inner edge for a H$_2$-rich atmosphere.
}
\label{fig:spectral}
\end{figure*}

% --
It is not obvious that background gases like H$_2$ or N$_2$ should
have a strong impact near the inner edge of the habitable
zone. Earth's greenhouse effect is already dominated by H$_2$O, and
H$_2$O becomes even more important at higher temperatures.
We therefore consider the inner edge for four atmospheric scenarios.
%--- 
First, we consider a pure H$_2$O atmosphere as the
simplest model of the runaway greenhouse
\citep{pierrehumbert2010,goldblatt2015a}. Second, we consider an Earth-like
N$_2$-H$_2$O atmosphere with 1 bar of N$_2$, which is typical of
conventional habitable zone calculations \citep{kasting1993,kopparapu2013b}.  Third, we consider the extreme
case of a CO$_2$-H$_2$O atmosphere with 1 bar of CO$_2$. Usually
CO$_2$ concentrations are assumed to be negligibly low near the inner edge of the
habitable zone due to the silicate-weathering thermostat
\citep{walker1981}. Here we include CO$_2$ as a limiting case of a
high MMW background gas with a strong greenhouse effect to draw
out the distinction between high MMW gases and H$_2$.
Fourth, we consider the new limit of a H$_2$-H$_2$O atmosphere with 1
bar of H$_2$.

% -- discuss albedo
Figure \ref{fig:spectral}a shows that in all four scenarios albedo
first decreases and then rises again with warming. This behavior is
due to the increase of atmospheric water vapor with warming.
Below about 250 K all atmospheres contain little water vapor, and are
dominated by atmospheric scattering and surface absorption.  At these
temperatures H$_2$ creates a significantly higher albedo than both
N$_2$ and CO$_2$, because H$_2$ has the highest scattering efficiency
among common gases on a per-mass basis (a kg of H$_2$ contains
more molecules than any other gas)\footnote{Note that our
    calculations compare different atmospheric compositions at
    fixed surface pressure, which is equivalent to holding the
    total atmospheric mass fixed.}.
Above 250 K shortwave absorption increases in the
near-IR, where H$_2$O is a strong absorber, so albedo drops.
Above 350 K the atmospheric mass starts to
increase in all scenarios, which increases Rayleigh scattering at short
wavelengths where H$_2$O is a relatively poor absorber, so albedo rises again.
Finally, above 400 K, all four scenarios
become dominated by water vapor so the albedo converges.

% -- discuss OLR
Figure \ref{fig:spectral}b shows that OLR in all scenarios approaches
a single limiting value at high temperatures.
The limiting value at high temperatures is called the runaway
greenhouse or Simpson-Nakajima limit \citep{goldblatt2013}. The
runaway arises from the rapid increase of water vapor with warming:
once water vapor becomes optically thick at all frequencies, longwave
emission decouples from the surface and OLR is determined by the
temperature of the upper atmosphere, which is constant once the
atmosphere becomes dominated by water vapor
\citep{nakajima1992,pierrehumbert2010,goldblatt2013}.

The behavior of OLR leading up to the runaway, however, depends strongly on
the assumed background gas. The simplest case is a pure H$_2$O
atmosphere, in which OLR approaches the runaway greenhouse
monotonically.
Adding 1 bar of N$_2$ does not change OLR much
relative to a pure H$_2$O atmosphere, because N$_2$ has no strong
greenhouse effect of its own, but it introduces an important
qualitative difference: N$_2$ causes OLR to
first overshoot the runaway limit before it slowly falls back \citep{nakajima1992}.
Adding 1 bar of CO$_2$ reduces OLR at low temperatures, due to
CO$_2$'s greenhouse effect, but at high temperatures CO$_2$ again
causes OLR to overshoot the runaway limit.
Previous work argued that the overshoot occurs because adding a dry
background gas steepens the atmospheric lapse rate relative to that of a steam atmosphere
\citep[in the sense of increasing $dT/dp$;][]{nakajima1992,pierrehumbert2010}. This means the upper
atmosphere is colder for a N$_2$-H$_2$O or CO$_2$-H$_2$O atmosphere than for pure
H$_2$O, which reduces the amount of water vapor in the upper
atmosphere and allows the atmosphere to emit more
radiation.
Importantly, this explanation seems insensitive to the composition of
the background gas, so one might expect that H$_2$ also creates
an OLR overshoot.

Surprisingly, we find that OLR changes very differently with warming
in a H$_2$-rich atmosphere (Fig.~\ref{fig:spectral}b). At cold temperatures H$_2$ reduces OLR
due to its CIA greenhouse effect. The change in OLR is
much larger for 1 bar of H$_2$ than for 1 bar of CO$_2$, so on a
per-mass basis H$_2$ is a stronger greenhouse gas than CO$_2$.
Above $250$ K OLR shoots up, however, then abruptly flattens out at $320$
K, before slowly approaching the runaway limit.
Crucially, whereas N$_2$ and CO$_2$ cause OLR to overshoot, H$_2$
causes OLR to undershoot the runaway greenhouse limit. This
undershoot strongly affects the climates of H$_2$-rich
atmospheres near the inner edge of the habitable zone.

% -- discuss stability
Figure \ref{fig:spectral}c combines our albedo and OLR calculations, and
shows how different background gases influence a planet's climate
stability.
We find that N$_2$-rich and CO$_2$-rich atmospheres abruptly cease to be stable above
surface temperatures of 310 K and 380 K. There are no nearby stable
states, so a N$_2$-rich or CO$_2$-rich planet transitions from a
moderately warm climate straight into the runaway greenhouse and would
only re-equilibrate again once the surface temperature exceeds about
$1,600$ K \citep{goldblatt2013}.
The abrupt onset of the runaway greenhouse in N$_2$-rich and CO$_2$-rich atmospheres
is largely due to the overshoot in the OLR-$T_s$
curve (Fig.~\ref{fig:spectral}b): once peak OLR is reached, the planet
cannot emit more longwave radiation with warming, so the only way for
the climate to remain stable is by reflecting more sunlight with
further warming. Albedo does increase slightly above $350$ K, due to increased Rayleigh scattering
as the atmosphere thickens, but the albedo increase is small and largely
cancelled by the decrease in OLR with further warming at these high
temperatures\footnote{Our N$_2$ calculations do show a small island
  of stability corresponding to surface temperatures of $390-410$ K
  and stellar fluxes of about $1356-1358$ W m$^{-2}$. It is highly
  unlikely that these states would actually be stable:
  our code still underestimates shortwave absorption in the near UV,
  and thus produces higher-than-realistic albedos, which would further
  destabilize warm climates. Moreover, the incident stellar flux would
  have to be extremely fine-tuned to maintain this state.}.

In contrast to high MMW atmospheres, pure H$_2$O and H$_2$-rich
atmospheres enter the runaway gradually which allows them to
  sustain unusually warm climates near the inner edge.
First, the inner edge of the habitable zone moves slightly
  outward in pure H$_2$O and H$_2$-rich  atmospheres when compared
  to high MMW atmospheres due to a the lack of an OLR overshoot. A
  high MMW background thus allows a
  planet to remain habitable at moderately higher stellar fluxes than is
  possible with a pure H$_2$O or a H$_2$-rich atmosphere.
  Focusing on those climates that are inside the habitable zone,
  however, we find that pure H$_2$O and H$_2$-rich atmospheres can
  sustain much higher surface temperatures than high MMW atmospheres.
  For example, an atmosphere with 1 bar of N$_2$ is essentially
  limited to surface temperatures below 310 K because higher surface
  temperatures are unstable and lead to further warming. In
  comparison, an atmosphere with 1 bar of H$_2$ does not lose its
  climate stability as it approaches the inner edge, and can
  sustain surface temperatures of $340$ K or more without entering the
  runaway.
This gradual approach to the runaway arises
from the OLR undershoot in H$_2$ (Fig.~\ref{fig:spectral}).

%% ----
\begin{figure}[b]
\centering
\includegraphics[width=\linewidth,clip]{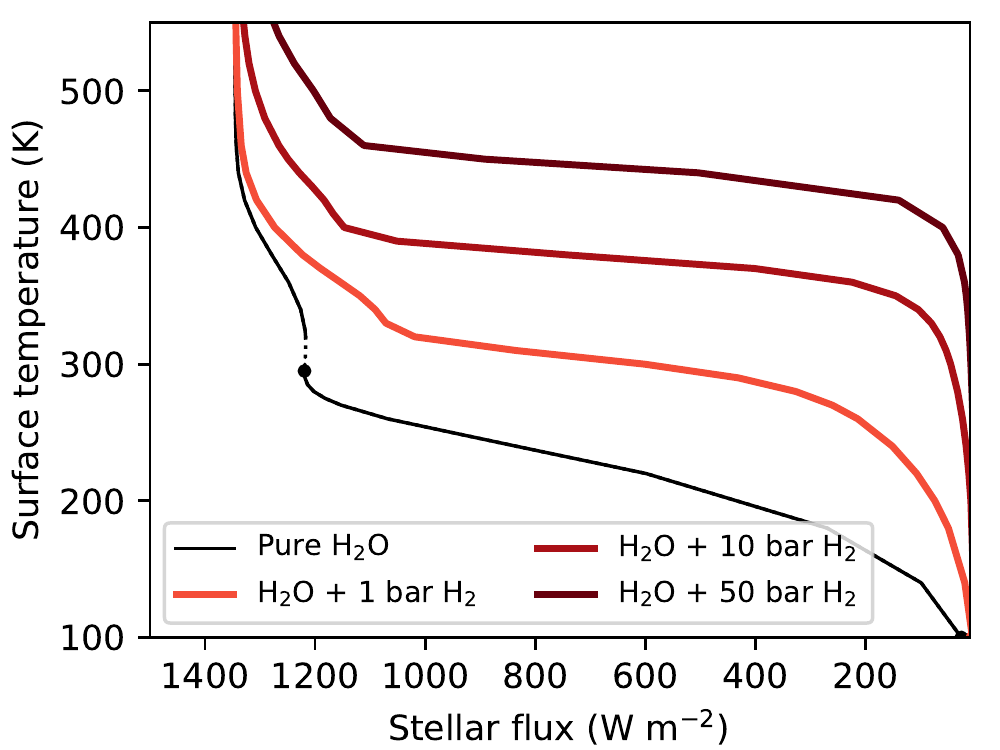}
\caption{The effect of H$_2$ at higher surface pressures. 50 bars of
  H$_2$ can warm the surface up to 450 K or more. The
  critical point of H$_2$O occurs at 647 K. Below the critical point a
  planet can maintain a liquid surface ocean, and is thus still
  potentially habitable.}
\label{fig:spectral_highps}
\end{figure}

  To explore how the warming effect of H$_2$ depends on surface
  pressure, we performed additional simulations in which we increased
  the surface pressure of H$_2$ up to 50 bars.
  Figure \ref{fig:spectral_highps} shows that 50 bars of H$_2$ are
  sufficient to warm the surface to 450 K or more near the inner edge
  of the habitable zone.
  For reference, the critical point of H$_2$O occurs at about 647
  K. As long as the surface temperature remains below the critical
  point, a planet can still maintain an ocean and thus remains habitable
  according to the classical definition of the habitable zone
  \citep{kasting1993}.
  Surface temperatures of 450 K or more, sustained over a timescale of
  100 Myrs, would strongly shape a planet's earliest surface
  environment and the development of prebiotic chemistry. Moreover, although these
  temperatures might seem high, the reported upper limit for extant life on Earth is around 400 K
  \citep{takai2008}, so these conditions could still be compatible
  with a thermophile biosphere.

\begin{figure}
\centering
\includegraphics[width=\linewidth, clip]{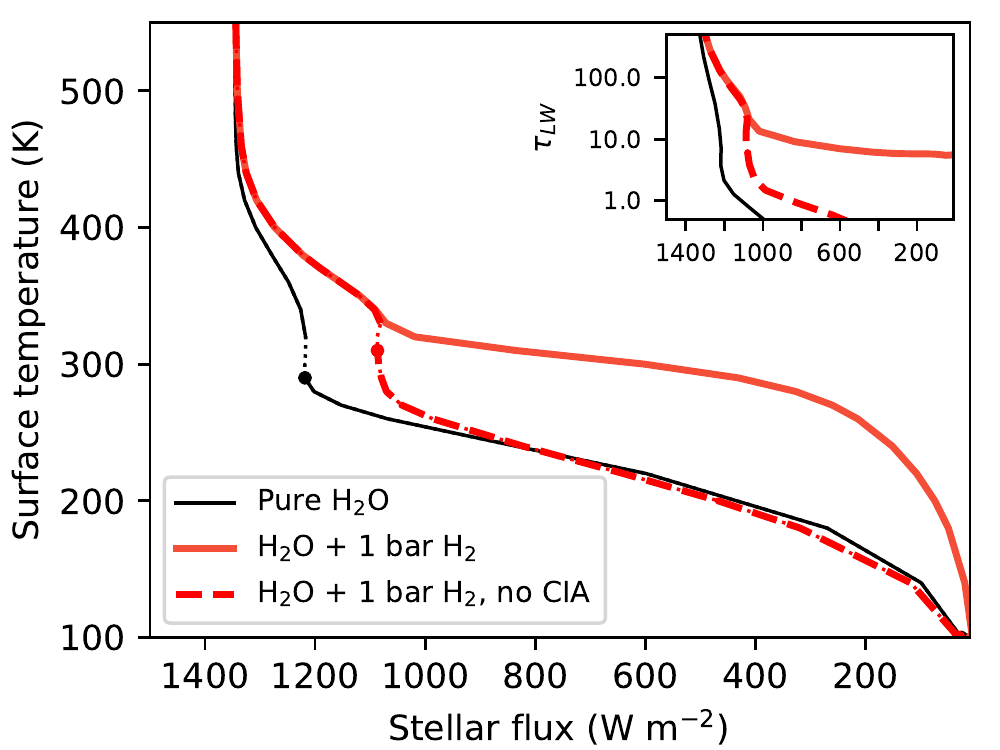}
\caption{The effect of H$_2$ near the inner edge is driven by H$_2$'s
  thermodynamics. The solid red line is our default calculation with
  H$_2$ infrared absorption (via H$_2$-H$_2$ collision-induced
  absorption; CIA), the dashed red line is a test in which we disable
  H$_2$ CIA in the longwave. The main plot shows surface temperature,
  the inset shows optical thickness as a function of stellar
  flux. Near the inner edge temperature and optical
    thickness are identical, so H$_2$'s impact near the inner edge
  is due to its thermodynamic greenhouse instead of its
  CIA greenhouse effect.}
\label{fig:spectral02}
\end{figure}

% ---
One might think that the strong H$_2$ warming near the inner edge is
caused by H$_2$'s radiative properties.
That is not the case. Figure \ref{fig:spectral02} shows that H$_2$
still leads to strong warming at surface temperatures above 320
K when we disable H$_2$-H$_2$ CIA in our OLR
calculations.
The inset in Figure \ref{fig:spectral02} shows a measure of the
  atmosphere's broadband longwave optical thickness, which we define
  as $\tau_{LW}=\ln[(\sigma T_s^4)^{-1}\times \int B_\lambda(T_s)
  e^{-\tau_\lambda} d\lambda]$, where $B_\lambda$ is the Planck
  function and $\tau_\lambda$ is the atmospheric optical thickness at
  a given wavelength $\lambda$. The inset shows that the optical
  thickness of a H$_2$-rich atmosphere is due to H$_2$'s CIA opacity below 320 K, but
  becomes dominated by H$_2$O opacity at high temperatures.
At low temperatures the climate effect of H$_2$ is therefore
due to its role as a greenhouse gas, but at high temperatures its
impact must be due to other physics.

%%%%%%%%%%%%%%%%%%%%%%%%%%%%
\subsection{An intuitive explanation for why MMW matters}
\label{sec:intuitive}

Why does H$_2$ lead to strong warming near the
inner edge, even without any greenhouse effect of its own, whereas a
heavy background gas like N$_2$ does not? We show next that
background gases influence OLR through their influence on an
atmosphere's temperature structure, which we quantify in terms of the
thermal scale height.  In turn, H$_2$'s ability to inflate the thermal
scale height is unique among common atmospheric gases.

First, the optical thickness of an atmospheric column is
directly proportional to the scale height of water vapor. To show this
we write the atmosphere's optical thickness\footnote{Although we will
assume grey radiation in the next section, the derivation in this
section is general and also holds monochromatically, i.e., at
a single wavelength.} as
\begin{eqnarray}
  \tau & = & \int_0^\infty \kappa_v \rho_v dz,
\end{eqnarray}
where $z$ is height, $\kappa_v$ is the absorption cross-section,
and $\rho_v$ is the density of water vapor (in general, the subscript
\textit{v} will denote quantities related to water vapor).
We assume $\kappa_v$ is independent of height and $\rho_v$ decays exponentially following
the water vapor scale height $H_{v}$, $\rho_v=\rho_{v,0} e^{-z/H_v}$.
In general $H_v$ can also vary with height, but due
to the exponentially rapid falloff of $\rho_v$ we 
treat $H_{v}$ as constant and equal to its value at the
surface. We additionally define the absorption length
$l_0\equiv 1/(\kappa_v \rho_{v,0})$, which is the characteristic
distance over which a longwave photon is absorbed by the atmosphere.
The atmospheric optical thickness is then simply
\begin{eqnarray}
  \tau & = & \kappa_v \rho_{v,0} \int_0^\infty e^{-z/H_v} dz,\\
       & = & \frac{H_v}{l_0}.
\label{eqn:tau}
\end{eqnarray}
The absorption length $l_0$ does not depend on background gases, so
N$_2$ or H$_2$ affect OLR only via changes in the scale height of water vapor $H_v$.

Next, background gases affect $H_v$ only via changes in the temperature scale height $H_T$. To show this we start with
the general definition for the scale height of water vapor,
\begin{eqnarray}
  H_v & = & \left( \frac{-1}{\rho_v} \frac{d\rho_v}{dz} \right)^{-1}.
\end{eqnarray}
% ---
We evaluate $H_v$ using the ideal gas law for water vapor, $\rho_v =
e^*/(R_v T)$ where $e^*$ is the saturation vapor pressure, and the
Clausius-Clapeyron relation, $d\ln e^*/d\ln T=L_v/(R_v T)$ where $L_v$
is the latent heat of vaporization,
\begin{eqnarray}
  \frac{1}{\rho_v} \frac{d\rho_v}{dz} & = & \frac{1}{e^*}
                                            \frac{de^*}{dz} - \frac{1}{T}\frac{dT}{dz},\\
      & = & \left(\frac{1}{e^*}\frac{de^*}{dT} - \frac{1}{T}\right)\frac{dT}{dz},\\
      & = & \left(\frac{L_v}{R_v T} - 1\right)\frac{1}{T}\frac{dT}{dz},\\
      & \approx & \frac{L_v}{R_v T} \frac{1}{T}\frac{dT}{dz}.
\end{eqnarray}
Here we used $L_v/(R_v T) \gg 1$, which holds for water vapor and a
wide range of other condensible gases. We recognize $(-1/T\times dT/dz)^{-1}$
as the temperature scale height $H_T$, which sets the vertical length
scale of temperature decrease, so
\begin{eqnarray}
  H_v & = & \frac{R_v T}{L_v} H_T.
\end{eqnarray}
Again, $R_v$, $T$ and $L_v$ are independent of any background gases,
so N$_2$ or H$_2$ affect optical thickness only via changes in $H_T$.

Next, we constrain the value of $H_T$. At cold temperatures water
vapor is highly dilute and has essentially no impact on the vertical
structure of the atmosphere. In this limit the lapse rate follows the
standard dry adiabat, $dT/dz = -g/c_p$, and
\begin{eqnarray}
  H_T^{dilute} & = & \frac{c_p T}{g}.
\label{eqn:Ht_dilute}
\end{eqnarray}
Here $c_p$ is the specific heat capacity of the background gas, e.g.,
of N$_2$.

At high temperatures the atmosphere becomes steam-dominated,
and total pressure becomes equal to the saturation vapor
pressure of water. In this limit the temperature scale height is
equal to
\begin{eqnarray}
  H_T^{steam} & = & -T\frac{dz}{dT}, \\
             & = & -T\frac{dz}{de^*} \frac{de^*}{dT}, \\
              & = & T\frac{1}{\rho_v g} \frac{L_v e^*}{R_v T^2}, \\
              & = & \frac{L_v}{g},
\label{eqn:Ht_steam}
\end{eqnarray}
where we used the hydrostatic relation in the third step.

If we evaluate $H_T$ for H$_2$ versus N$_2$, we find that different background
gases lead to dramatically different temperature scale heights.
With an Earth-like gravity and at a temperature of 300 K, $H_{T}^{dilute}
 \approx 30$ km in N$_2$, whereas $H_{T}^{dilute}
 \approx 430$ km in H$_2$.
The difference arises from the low molecular weight of H$_2$ relative
to that of N$_2$. Even though both are diatomic molecules with similar
degrees of freedom, one kilogram of H$_2$ contains 14 times more
molecules than one kilogram of N$_2$, which means the specific heat
capacity $c_p$ of H$_2$ is roughly 14 times larger than that of
N$_2$.

Now, as both N$_2$ and H$_2$ atmospheres warm up, they eventually end up
being dominated by water vapor. The temperature scale
height of a steam atmosphere is $H_{T}^{steam} \approx 250$ km. This
means $H_T$ grows with warming if an atmosphere starts out N$_2$-rich,
but $H_T$ shrinks with warming if the atmosphere starts out
H$_2$-rich.

Because $H_T$ controls $H_v$, and therefore the atmosphere's optical
thickness $\tau$, we see that warming leads to a compositional climate
feedback which depends on the molecular weight of the background gas.
An N$_2$-rich atmosphere starts out dense and compact, with a small
thermal scale height $H_T$. Warming moves the atmospheric composition closer
to the steam limit, so $H_T$ increases. This implies $H_v$ also
increases with warming, so moving from a N$_2$-rich to a H$_2$O-rich
atmosphere means the atmosphere moistens even faster than a
pure H$_2$O atmosphere would. The moistening increases $\tau$, tends
to reduce OLR, and thus acts to amplify warming.

In contrast, a H$_2$-rich atmosphere starts out with a large
scale height $H_T$. Warming moves the atmospheric composition closer to
the steam limit, so $H_T$ and $H_v$ eventually have to decrease.
Consequently, the shift from H$_2$ to H$_2$O means the atmosphere
moistens less quickly than a pure H$_2$O or a N$_2$-rich atmosphere
does -- a stabilizing climate feedback.

The existence of this feedback, and its sensitivity to the MMW of the
background gas, provide a first clue as to why H$_2$-rich
atmospheres become much warmer than N$_2$-rich, or even CO$_2$-rich, atmospheres
(Fig.~\ref{fig:spectral}).
Moreover, the sign of this feedback appears to be stabilizing only for
H$_2$. Even if we consider Helium as the next-heavier background gas
after H$_2$, such as a remnant atmosphere clinging to the core of a
evaporated mini-Neptune \citep{hu2015}, we find
$H_{T}^{dilute} \approx 160$ km, which is smaller than
$H_{T}^{steam} \approx 250$ km.
The compositional feedback therefore requires the
molecular weight of the background gas to be small enough so that the
sensible heat of the background gas, $c_p T$, becomes
larger than the latent heat of steam, $L_v$.
H$_2$ appears to be unique in this regard, which means H$_2$-rich
habitable atmospheres are distinct in terms of
generating a stabilizing feedback as they transition from a cold, dry
climate to a hot, steam-dominated climate.
%

%%%%%%%%%%%%%%%%%%%%%%%%%%%%
\section{New climate states in H$_2$-rich atmospheres} % xxx 
\label{sec:grey}

Different background gases have a strong influence on a planet's
climate, so can we understand their effect more generally?
In this section we first present grey radiative calculations
which qualitatively recover the onset of the runaway greenhouse in our spectral
  calculations, and which therefore allow us to understand in detail
  how different background gases affect the inner edge of the
  habitable zone (Section \ref{sec:grey_olr}).
Second, we present analytical results
which explain why OLR overshoots the runaway limit with a high MMW gas
like N$_2$, but undershoots the runaway limit with H$_2$. In doing so
we present a new climate state which we call the dilute
runaway, and elucidate its underlying physics (Sections \ref{sec:dilute_runaway} and
\ref{sec:dilute_runaway_conditions}).
Third, we discuss how H$_2$'s thermodynamics give rise
  to a non-monotonic climate feedback which we call the Souffl\'e effect (Section \ref{sec:souffle}).
  Finally, the calculations in this section assume idealized
  grey radiation, which allows us to understand the
  thermodynamic effect of different background gases, but do not
  capture all aspects of our spectral calculations. We discuss the
  remaining differences between grey and spectral calculations
  at the end of this section (Section \ref{sec:dry_absorber}).

\begin{figure*}
\centering
\includegraphics[width=0.75\linewidth,clip]{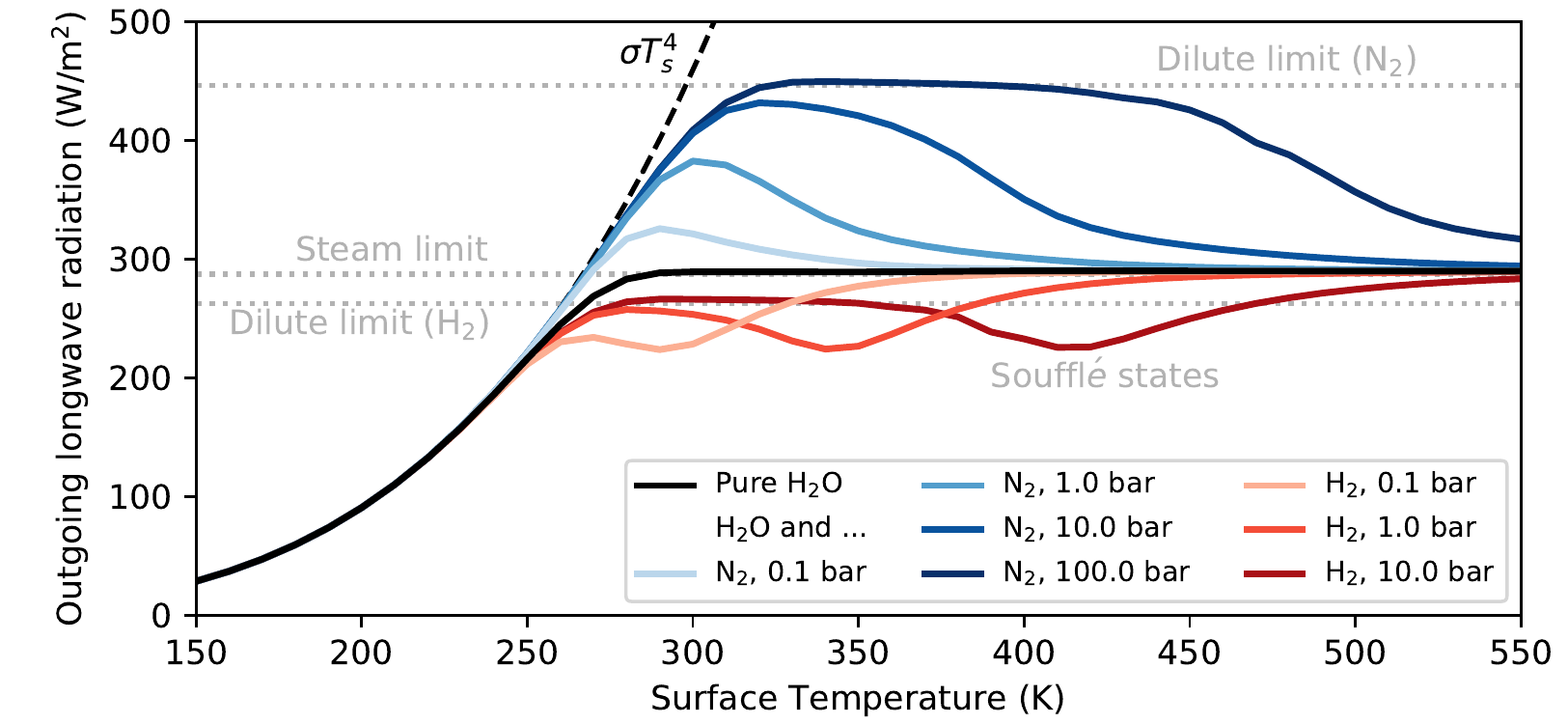}
\caption{Background gases can have a strong influence on a planet's
  outgoing longwave radiation (OLR).
  Colors show grey calculations with N$_2$ and H$_2$ backgrounds,
  grey shows analytical results which we derive in Section
  \ref{sec:dilute_runaway}. OLR tends
  towards the steam limit at high temperatures, but at moderate
  temperatures and high surface pressures we find an additional
  ``dilute runaway'' limit. The dilute limit in N$_2$ is
  greater than the steam limit, so N$_2$ atmospheres overshoot the
  steam limit, whereas in H$_2$ it is less than the steam limit, so
  H$_2$ atmospheres undershoot. In addition, the transition from the dilute limit to
  the steam limit in H$_2$ is non-monotonic due to the ``Souffl\'e effect''
  (Sec.~\ref{sec:souffle}). }
\label{fig:grey}
\end{figure*}

%%%%%%%%%%%%%%%%%%%%%%%%%%%%
\subsection{Numerical calculations}
\label{sec:grey_olr}

For our grey radiative calculations we use the same grey longwave
opacity for water vapor as in \cite{nakajima1992},
$\kappa_v=0.01$ m$^2$ kg$^{-1}$. The background gas is assumed to be
completely transparent to longwave radiation, so its only influence on
OLR is via its influence on the atmosphere's temperature structure.
The colored lines in Figure \ref{fig:grey} show the response of OLR to
warming for three different atmospheric compositions: a pure H$_2$O
atmosphere, a N$_2$-rich atmosphere, and a H$_2$-rich atmosphere. We
vary the surface pressure of the background gas between 0.1 and 10 bar
for H$_2$ and 0.1 and 100 bar for N$_2$.

Figure \ref{fig:grey} recovers the main qualitative features of our
spectral calculations. First, the OLR of a grey pure H$_2$O atmosphere
closely resembles our spectral calculations for pure H$_2$O
(Fig.~\ref{fig:spectral}). At temperatures below about 250 K there is
little water vapor in the atmosphere, the atmosphere is largely
transparent to longwave radiation, and most emission originates from the
surface, so $\mathrm{OLR} \approx \sigma T_s^4$. With further warming
the atmosphere rapidly becomes optically thick and monotonically
approaches a limiting value of about $290$ W m$^{-2}$.
This limiting OLR has previously been called the Simpson-Nakajima limit
\citep{goldblatt2013}, for reasons we explain below we will call it
here the ``steam limit''.
Second, adding N$_2$ as a background gas leads to an overshoot and allows 
OLR to exceed the steam limit, as in our spectral calculations.
Adding H$_2$ has a similar but opposite effect, in which OLR
undershoots the steam limit.
Figure \ref{fig:grey} also shows several new features which we did not find
in our spectral calculations.
Although some of these features are sensitive to the details of
  our radiative assumptions (see Section \ref{sec:dry_absorber}), we
  focus on them here because they allow for important insight
  into the thermodynamic effect of different background gases.
In particular, at moderate surface temperatures and high surface
pressures, OLR hits an asymptotic limit that is distinct from the steam limit.
This new OLR limit is higher than the steam limit for
N$_2$ and lower for H$_2$. The temperature range over which OLR
remains close to this limit depends on the amount of background
gas. For example, atmospheres with less than 1 bar of background gas
do not reach this limit whereas an atmosphere with 100 bar of N$_2$ remains at
this limit for more than 100 K.
We note that this asymptotic limit is distinct from the
Komabayashi-Ingersoll limit \citep{komabayasi1967,ingersoll1969},
which requires a stratosphere, whereas our calculations do not include
a stratosphere (we discuss this point in Section \ref{sec:discussion}).

The asymptotic OLR limit at high surface pressures was first discussed
by \citet{nakajima1992}.
Imagine a moderately warm planet with large amounts of dry background
gas, so that its atmosphere contains some H$_2$O but water vapor does
not yet affect the temperature structure and the lapse rate is still dry-adiabatic. As the
planet warms beyond a threshold (e.g., $\sim$ 300 K in our spectral
calculations; Fig.~\ref{fig:spectral}), water vapor becomes optically thick and the
planet's OLR decouples from the surface temperature. In this case the
atmosphere's emission temperature remains constant with warming -- the
planet is in a runaway state.
% ---
The invariance to warming is only disrupted once water vapor reaches
a sufficient abundance to affect the lapse rate, and thus modify the temperature at the
emission level. Because this radiation
limit resembles the runaway greenhouse, but requires enough
background gas so that H$_2$O is dilute compared to the background gas,
we call this OLR limit the ``dilute runaway''. The dilute runaway stands in contrast to
the traditional Simpson-Nakajima runaway in which the atmosphere
is dominated by water vapor, and which we call the ``steam runaway''.

\citet{nakajima1992} discussed the physics that can lead to the dilute runaway, but
did not discuss its physical relevance in detail. Below we
show that the dilute runaway, and how it depends on atmospheric MMW,
is the underlying reason why H$_2$-rich atmospheres approach the
steam runaway gradually, whereas N$_2$-rich
atmospheres are unstable at high temperatures and enter the steam
runaway abruptly.

%%%%%%%%%%%%%%%%%%%%%%%%%%%%
\subsection{The dilute runaway}
\label{sec:dilute_runaway}

As an atmosphere warms its OLR first tends towards the dilute limit,
and then towards the steam limit (Fig.~\ref{fig:grey}).
If one could derive a simple expression for OLR in both limits,
one could therefore predict whether a background gas causes an
atmosphere to over- or undershoot the steam limit.
In this section we do so, by deriving a closed-form analytical expression for the OLR of
a runaway greenhouse atmosphere which was previously only solved
implicitly \citep{nakajima1992,pierrehumbert2010}.

First, we derive an approximate form of the Clausius-Clapeyron
relation. The Clausius-Clapeyron relation is often written as $d\ln e^*/d\ln T=L_v/(R_v T)$, which
can be integrated to solve for $e^*(T)$. This form of $e^*(T)$ does not lead to closed-form
expressions for optical thickness and OLR, however, so we
instead treat the right-hand side as constant,
$d\ln e^*/d\ln T \approx L_v/(R_v T_0) \equiv \gamma$. In this case
$e^*$ becomes a simple power law,
\begin{eqnarray}
  e^* & = & e^*_0 \left(\frac{T}{T_0}\right)^\gamma.
\end{eqnarray}
Here $\gamma$ controls how rapidly saturation
vapor pressure increases with temperature, and $T_0$ is a reference temperature
which we will pick to be close to the temperature range of interest.

% -----------------
%
Our approximation allows us to express the relation
between optical thickness and temperature as a power law, which turns
out to be highly useful. We combine Equation \ref{eqn:tau} with the
ideal gas law for water vapor, $\rho_v = e^*/(R_v T)$. We find
\begin{eqnarray}
  \tau & = & \kappa_v H_v \rho_{v},\\
       & = & \kappa_v H_T \frac{e^*}{L_v},\\
       & = & \kappa_v H_T \frac{e_0^*}{L_v} \times \left( \frac{T}{T_0} \right)^\gamma.
\end{eqnarray}
This expression depends on atmospheric composition and
background gases only through the temperature scale height $H_T$. As we
showed before, $H_T^{dilute} = c_p T/g$ and $H_T^{steam} = L_v/g$. If
we further approximate the temperature dependency of $H_T^{dilute}$ as
 $H_T^{dilute} \approx c_p T_0/g$, we can express $\tau$ as single power law
 in both limits,
\begin{eqnarray}
  \tau & = & \tau_0 \left(\frac{T}{T_0}\right)^\gamma.
\end{eqnarray}
Here $\tau_0 = (c_p T_0/L_v) \times \kappa_v e_0^*/g$ in the dilute
limit while  $\tau_0 = \kappa_v e_0^*/g$ in the steam limit.

% ---
We are now able to find a closed solution for the runaway
greenhouse radiation limit. For a grey gas, OLR is equal to
\begin{eqnarray}
  \mathrm{OLR} & = & \sigma T_s^4 e^{-\tau} + \int_0^{\tau} \sigma T(\tau')^4 e^{-\tau'} d\tau'.
\end{eqnarray}
As the atmosphere becomes optically thick the surface contribution from
the first term can be neglected and the integral's upper limit replaced with infinity,
\begin{eqnarray}
  \mathrm{OLR}_\infty & \approx & \int_0^{\infty} \sigma T(\tau)^4 e^{-\tau} d\tau,\\
  & = & \sigma T_0^4 \int_0^{\infty} \left(\frac{\tau}{\tau_0}\right)^{4/\gamma} e^{-\tau} d\tau,\\
  & = & \sigma T_0^4 \tau_0^{-4/\gamma} \int_0^{\infty} \tau^{4/\gamma} e^{-\tau} d\tau,\\
        \mathrm{OLR}_\infty & = & \sigma T_0^4 \times \frac{ \Gamma\left(1+4/\gamma\right)}{\tau_0^{4/\gamma}}.
\label{eqn:OLR_inf}
\end{eqnarray}

Here $\Gamma$ is the gamma function, and $\tau_0$ is defined as 
\begin{eqnarray}
  \tau_0 & = & \begin{cases}
\frac{\kappa_v e^*(T_0)}{g}, ~~\mathrm{steam~limit}\\
\frac{\kappa_v e^*(T_0)}{g} \frac{c_p T_0}{L_v}, ~~\mathrm{dilute~limit}.
\end{cases}
\end{eqnarray}
Our solution in Equation \ref{eqn:OLR_inf} has a similar form to
\citet{pierrehumbert2010}'s solution for the OLR of a dry atmosphere,
but instead holds for a moist runaway atmosphere.

To evaluate our solution we only need to choose a reference
temperature $T_0$ where our approximation of 
Clausius-Clapeyron will be most accurate. For the runaway
greenhouse $T_0$ should be close to the surface temperature at which the
atmosphere becomes optically thick.
A convenient choice for all three atmospheric scenarios is the
temperature at which the optical thickness of a pure-H$_2$O atmosphere
equals unity.

Figure \ref{fig:grey} compares our analytical solution against our
numerical calculations. We find that the analytical solution closely match
the numerical results, so our expression successfully captures the
physics of both the dilute and the steam runaway.

Our solution now allows us to understand why OLR overshoots in a
N$_2$-rich atmosphere but undershoots in a H$_2$-rich atmosphere
(Fig.~\ref{fig:grey}).
The reference temperature $T_0$ is independent of the background gas,
so the greater value of $\mathrm{OLR}_\infty$ for N$_2$ relative to H$_2$
depends entirely on $\tau_0$. In turn, background gases only affect
$\tau$ through the temperature scale height $H_T$.
The ratio of optical thicknesses in the dilute
versus the steam limit is
\begin{eqnarray}
\frac{\tau_{\mathrm{dilute}}}{\tau_{\mathrm{steam}}} & = & \frac{H_T^{dilute}}{H_T^{steam}},\\
                                                           & = & \frac{c_p T_0}{L_v},
\label{eqn:tau_ratio}
\end{eqnarray}
which expresses our result from Section \ref{sec:intuitive} in terms
of a single non-dimensional parameter. Again, for H$_2$O condensing in
N$_2$ at around 300 K this ratio is less than unity,
$c_p T_0/L_v \sim 0.1$, whereas in H$_2$ this ratio is
$c_p T_0/L_v \sim 1.7$. Comparing the two values, the ratio is closer
to unity for H$_2$ than for N$_2$, which explains why the OLR undershoot in
H$_2$ is small whereas the OLR overshoot in N$_2$ is large
(Fig.~\ref{fig:grey}).  Moreover, as we showed in Section
\ref{sec:intuitive}, H$_2$'s impact as a background gas is essentially unique. Even if we
consider condensible gases other than H$_2$O, we are not able to find
a likely combination for which $c_p T_0/L_v > 1$ without H$_2$ as the
background gas\footnote{For CO$_2$ condensing in a 
  CH$_4$-rich atmosphere, $c_p T_0/L_v \sim 1.1$ due to CO$_2$'s small
  latent heat combined with CH$_4$'s relatively small MMW. However,
  such an atmosphere would likely not be chemically
  stable unless it was constantly replenished through a disequilibrium
  process.}. Conversely, even if we consider gases other than H$_2$O
that could condense in a H$_2$-rich atmosphere (e.g., an early Titan
with a H$_2$ atmosphere and a CH$_4$ surface ocean, or a super-Mars
with a H$_2$ atmosphere and large CO$_2$ glaciers), we find that $c_p
T_0/L_v > 1$ for all these condensible substances.

In conclusion, H$_2$ is the only common background gas for which the dilute
runaway limit lies below the steam limit, which explains why H$_2$-rich
atmospheres approach the runaway greenhouse gradually while all high
MMW atmospheres enter the runaway greenhouse abruptly. Next, we
show why the evolution of OLR is not monotonic with warming in a
H$_2$ atmosphere.

%%%%%%%%%%%%%%%%%%%%%%%%%%%%
\subsection{Non-monotonic OLR: the Souffl\'e Effect in H$_2$}
\label{sec:souffle}

\begin{figure*}
\centering
\includegraphics[width=\linewidth, clip]{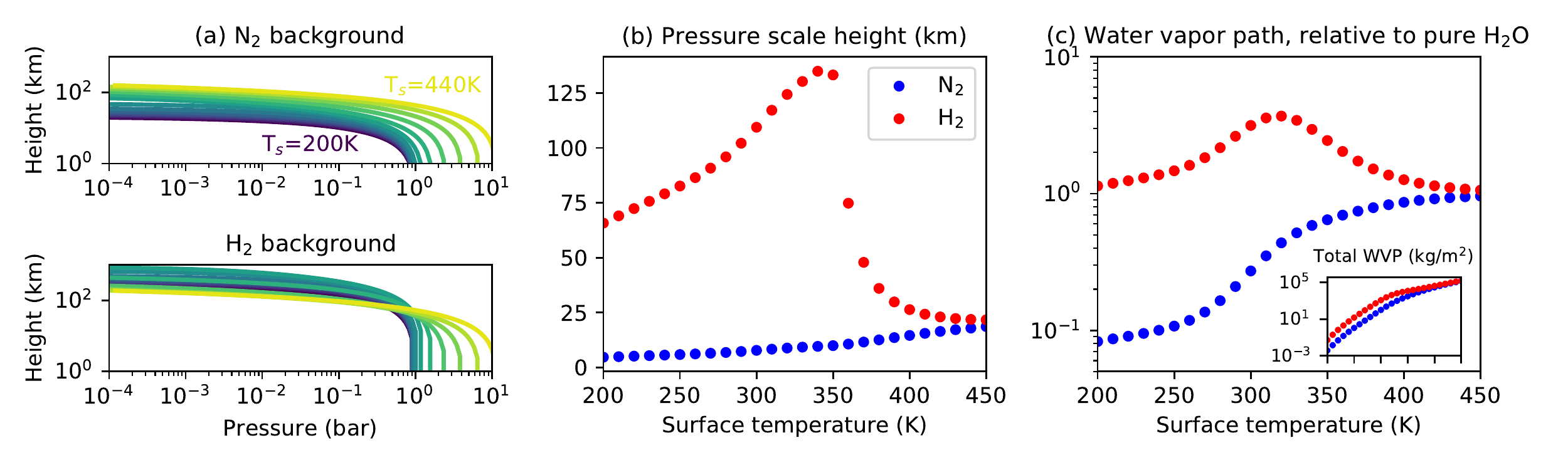}
\caption{The Souffl\'e effect: in a H$_2$-rich atmosphere, moderate
  warming increases the atmospheric water vapor, which releases latent
  heat and puffs the atmosphere up further. However, strong warming
  collapses the atmosphere, once heavy water vapor and condensation
  start to dominate over light H$_2$ and dry convection. In contrast, N$_2$-rich
  atmospheres puff up monotonically. a) Pressure-height profiles as a
  function of surface temperature. b) Average pressure scale height,
  $p_s^{-1} \int (-1/p\times dp/dz)^{-1} dp$, versus surface
  temperature. c) The atmospheric water vapor path (WVP) as function
  of surface temperature, relative to the WVP of a pure H$_2$O
  atmosphere. Inset shows absolute values of WVP.}
\label{fig:scale_height}
\end{figure*}

In the previous section we derived asymptotic expressions for the dilute
and steam limits. In this section we go one step further and consider
first-order effects that appear when an atmosphere is slightly warmer
than the cold dilute limit, or slightly colder than the hot steam
limit.
We find two thermodynamic feedbacks that shape the evolution of OLR
with warming, and that explain the non-monotonicity of OLR in H$_2$ (Fig.~\ref{fig:grey}).

At relatively low temperatures, we find that the latent heat release of water vapor
dominates over water vapor's other thermodynamic effects.
This means the addition of water vapor will always first warm and moisten an atmosphere,
regardless of the dry background gas. As proof we consider the general
moist adiabat \citep{pierrehumbert2010,ding2016} in the limit of a
nearly dry atmosphere. We perform a series expansion, assuming the molar
mixing ratio is small, $e^*/p_d \ll 1$, and additionally use $L_v/(R_v
T) \gg 1$ (which holds for a wide range of condensibles) while $R/c_p$ is always of order unity. We find
\begin{eqnarray}
    \left.\frac{d T}{d p}\right|_{dilute} & \approx & \frac{R_d}{c_p} \frac{T}{p} \times
          \left( 1 -  \frac{R_d}{c_p} \left(\frac{L_v}{R_v T}\right)^2 \frac{e^*}{p}   \right).
\label{eqn:dilute_first_order}
\end{eqnarray}
By combining this result with the hydrostatic relation for dry air we
can solve for the thermal scale height in the near-dilute limit,
\begin{eqnarray}
  H_T & = & -T \frac{dz}{dp} \frac{dp}{dT},\\
      & = & \frac{c_p T}{g} \left( 1 + \frac{R}{c_p}\left(\frac{L_v}{R_v T}\right)^2\frac{e^*}{p}\right).
\label{eqn:tau_dilute_first_order}
\end{eqnarray}

Here the factor $R_d/c_p \times (L/(R_v T))^2 \times e^*/p$ captures
the first-order thermodynamic effect of water vapor. Because
$R_d/c_p \times (L/(R_v T))^2 \times e^*/p$ is always positive, it
acts to increase $H_T$ and thus $\tau$. Intuitively, adding trace
amounts of water vapor to a dry parcel releases latent heat, which
decreases the lapse rate and allows the atmospheric column to hold
more water vapor. This provides a moistening feedback which increases
$H_v$ and $\tau$, and thus decreases OLR.  Importantly,
the moistening feedback is largely independent of background gas
because the background gas only appears via the parameter $R_d/c_p$
which is approximately constant (e.g., $R_d/c_p = 2/5$ for He, whereas
$R_d/c_p \approx 2/7$ for N$_2$ or H$_2$). This feedback explains why,
after an atmosphere hits the dilute runaway, further warming then
leads to a OLR decrease in both N$_2$- and H$_2$-rich atmospheres
(Figure~\ref{fig:grey}).

At high temperatures the compositional feedback we already described
in Section \ref{sec:intuitive} dominates.
To quantify it we consider the effect of a background gas on the mean
molecular weight of a steam atmosphere, which shows up via the gas
constant used in the hydrostatic relation, and perform a series
expansion in the limit $p_d/e^* \ll 1$.
We find
 \begin{eqnarray}
   \left.\frac{dp}{dz}\right|_{steam} & = & -\bar{\rho} g,\\
   & \approx & \rho_v g \times \left( 1 + \frac{R_d-R_v}{R_d} \frac{p_d}{e^*} \right)
\label{eqn:steam_first_order}
 \end{eqnarray}
We solve for the thermal scale height by combining this result with
the lapse rate of a steam atmosphere and find
\begin{eqnarray}
H_T & = & \frac{L_v}{g} \left( 1 + \frac{R_d-R_v}{R_d} \frac{p_d}{e^*} \right).
\label{eqn:tau_steam_first_order}
\end{eqnarray}

The sign of $(R_d-R_v)/R_d \times p_d/e^*$ depends on the MMW of the background gas, so a
background gas can either increase or decrease
$H_T$ and $\tau$ near the steam limit. For N$_2$ in H$_2$O, $R_d <
R_v$, so
N$_2$ has a net drying effect. However, as a steam atmosphere warms,
this drying effect weakens and
$\tau$ will increase even faster than it would in a pure
H$_2$O atmosphere - a positive feedback, just as in Section
\ref{sec:intuitive}.
In contrast, H$_2$ has a net moistening effect because $R_v < R_d$.
This effect weakens with warming, which slows the increase of $\tau$
relative to that in a pure H$_2$O atmosphere, and thus provides a
negative feedback.

For a N$_2$-dominated atmosphere, both the latent heating feedback in
Equation \ref{eqn:tau_dilute_first_order} and the drying feedback in
Equation \ref{eqn:tau_steam_first_order} have the same sign with
warming and act to decrease OLR. That is why, after an initial overshoot, OLR decreases
monotonically towards the steam limit
(Figure~\ref{fig:grey}). Conversely, for a H$_2$-dominated atmosphere
the latent heating feedback first decreases OLR before the moistening
feedback eventually sets in and causes OLR to rise again.

%% ---
Based on what a H$_2$-rich atmosphere looks like as it warms up,
we call its temperature-dependent succession of feedbacks the
``Souffl\'e effect''.  Figure \ref{fig:scale_height} shows how the
Souffl\'e effect plays out in our calculations. As the atmosphere warms
up, the pressure scale height increases. At cold temperatures this
increase is linear with temperature, and arises from the atmosphere's
thermal expansion.
Starting at 280 K the H$_2$-rich atmosphere begins to puff up a lot
faster. This effect is due to the latent heat release of water vapor,
which moistens the atmosphere and reduces OLR
(Fig.~\ref{fig:grey}). However, just as too much steam causes a
Souffl\'e to collapse, too much steam becomes fatal for a
H$_2$-rich atmosphere: once heavy condensing H$_2$O begins to displace
light dry H$_2$, at around 340 K, the atmospheric scale
height collapses. This collapse slows down the further increase of
atmospheric water and increases OLR  (compare Fig.~\ref{fig:grey} and
Fig.~\ref{fig:scale_height}b).

%%%%%%%%%%%%%%%%%%%%%%%%%%%%
\subsection{How much dry background gas is required?}
\label{sec:dilute_runaway_conditions}

\begin{table}[b!]
%\centering
\begin{tabular}{ccc}
\hline
 & Dilute Runaway  & Steam Runaway \\
 Condensing gas & $p_{s,dry}$ greater than (bar) &  $p_{s,dry}$ less than (bar)\\
\hline
H$_2$O &  2.0 & 0.004  \\
CH$_4$ &  13 &  0.05   \\
NH$_3$ &  16 &  0.03   \\
CO$_2$ &  420 &  4.8     \\
\hline
\end{tabular}
\caption{An atmosphere's dry background pressure determines whether an atmosphere
  can enter a dilute runaway, or transitions directly into the
  steam runaway. For the dilute runaway we assume a diatomic
  background gas, $R_d/c_p$=2/7, for the steam runaway we use
  N$_2$ as the background gas.} \label{tab:necessary_conditions}
\end{table}

We have shown that different background gases can have a major
influence on a planet's climate, but how much background gas is required
for this influence to become apparent?
We address this question by considering whether an
atmosphere can transition straight from a cold climate into the steam
runaway, or whether it enters a dilute runaway phase first.

To enter either runaway state the atmosphere first needs to become optically
thick so that OLR decouples from the planet's surface emission, which
means
\begin{eqnarray}
\tau_{LW}(T_1)  & > & 1.
\end{eqnarray}
Here $T_1$ is the surface temperature necessary so that the overlying
atmospheric column contains enough water vapor to be optically
thick.
For example, with a grey absorption cross-section of $\kappa_v=0.01$ m$^2$ kg$^{-1}$, $T_1 \sim 310 $ K in
a thick N$_2$ background and $T_1 \sim 270 $ K in a thick H$_2$ background.
In general $T_1$ depends on the detailed radiative properties of the condensing gas. However,
it turns out that several condensible greenhouse gases start to become optically
thick roughly around surface temperatures corresponding to their triple point \citep[see SI
in][]{koll2018d}. In the following we therefore approximate $T_1$ as
the triple point temperature of each condensible gas\footnote{By
  equating $T_1$ with the triple point, our criterion is overly
  stringent for H$_2$O: H$_2$O's triple point is 273 K, whereas the
  H$_2$O runaway occurs at around 310 K in a saturated N$_2$-rich atmosphere
  \citep{koll2018d}.}.

To sustain the dilute runaway an atmosphere needs to have
sufficient background gas, otherwise it only experiences a
transient overshoot/undershoot (Fig.~\ref{fig:grey}).
We quantify the required amount of background gas by considering the
first-order correction term to the dry adiabat in Equation
\ref{eqn:dilute_first_order}, which is equal to $R_d/c_p \times (L_v/(R_vT))^2 \times
e^*/p$. For the atmosphere to be dilute this term
has to be much less than unity, which leads to the following threshold
on the dry (background) surface pressure,
\begin{eqnarray}
 p_{s,dry} & \gg & \frac{R_d}{c_p} \left(\frac{L_v}{R_v T_1}\right)^2 e^*(T_1).
\label{eqn:critical_dry_pressure}
\end{eqnarray}

% ---
Conversely, if the background pressure is very low the atmosphere
transitions from a cold climate straight into the steam
runaway. Equation \ref{eqn:steam_first_order} shows that the
first-order correction term near the steam limit is equal to
$(R_d-R_v)/R_d\times p_d/e^*$, which leads to the following threshold,
\begin{eqnarray}
 p_{s,dry} & \ll & \frac{R_d}{R_d-R_v} e^*(T_1).
\label{eqn:critical_dry_pressure02}
\end{eqnarray}
% ---
We evaluate the thresholds from Equations
\ref{eqn:critical_dry_pressure} and \ref{eqn:critical_dry_pressure02}
in Table \ref{tab:necessary_conditions}, replacing inequalities with a factor of three.

Table \ref{tab:necessary_conditions} shows that even small amounts of
dry background gas are sufficient to modify the behavior of a steam
atmosphere, while moderately large amounts of dry background gas
can sustain a dilute runaway.
A few millibar of dry background gas, or about the surface
pressure of Mars, already suffice to modify the H$_2$O steam runaway. This
value increases to several bar of dry gas for CO$_2$, while methane
and ammonia are intermediate between the two.
It is thus relatively easy for a minor background gas to modify the
OLR of a condensible-rich atmosphere. Of course, the exact climatic
impact of the background gas (i.e., whether it creates an OLR overshoot or undershoot)
depends on its MMW.
The dilute runaway is easiest to achieve for H$_2$O, with
around two bar of dry background gas, and most difficult to achieve
for CO$_2$, with several hundred bar of dry background gas
(Table \ref{tab:necessary_conditions}). The difference between
condensible gases is primarily driven by the triple point
vapor pressures, with a small vapor pressure for H$_2$O and a large
vapor pressure for CO$_2$.

To put these calculations into context, present-day Earth has 1 bar of
N$_2$-O$_2$ background while Titan has about 1.5 bar of N$_2$. In both cases
there is enough dry background gas to significantly affect the
atmosphere's thermal structure and OLR, but both fall short of the dilute
runaway threshold for H$_2$O and CH$_4$. 
In contrast, Venus' atmosphere currently has about 3 bar of N$_2$ background
pressure. If this N$_2$ entered the atmosphere early in
the planet's history, it could have had a significant influence on the
onset of the runaway greenhouse on Venus.

%%%%%%%%%%%%%%%%%%%%%%%%%%%%
\subsection{Explaining the difference between grey and spectral calculations}
\label{sec:dry_absorber}

\begin{figure}
\centering
\includegraphics[width=\linewidth,clip]{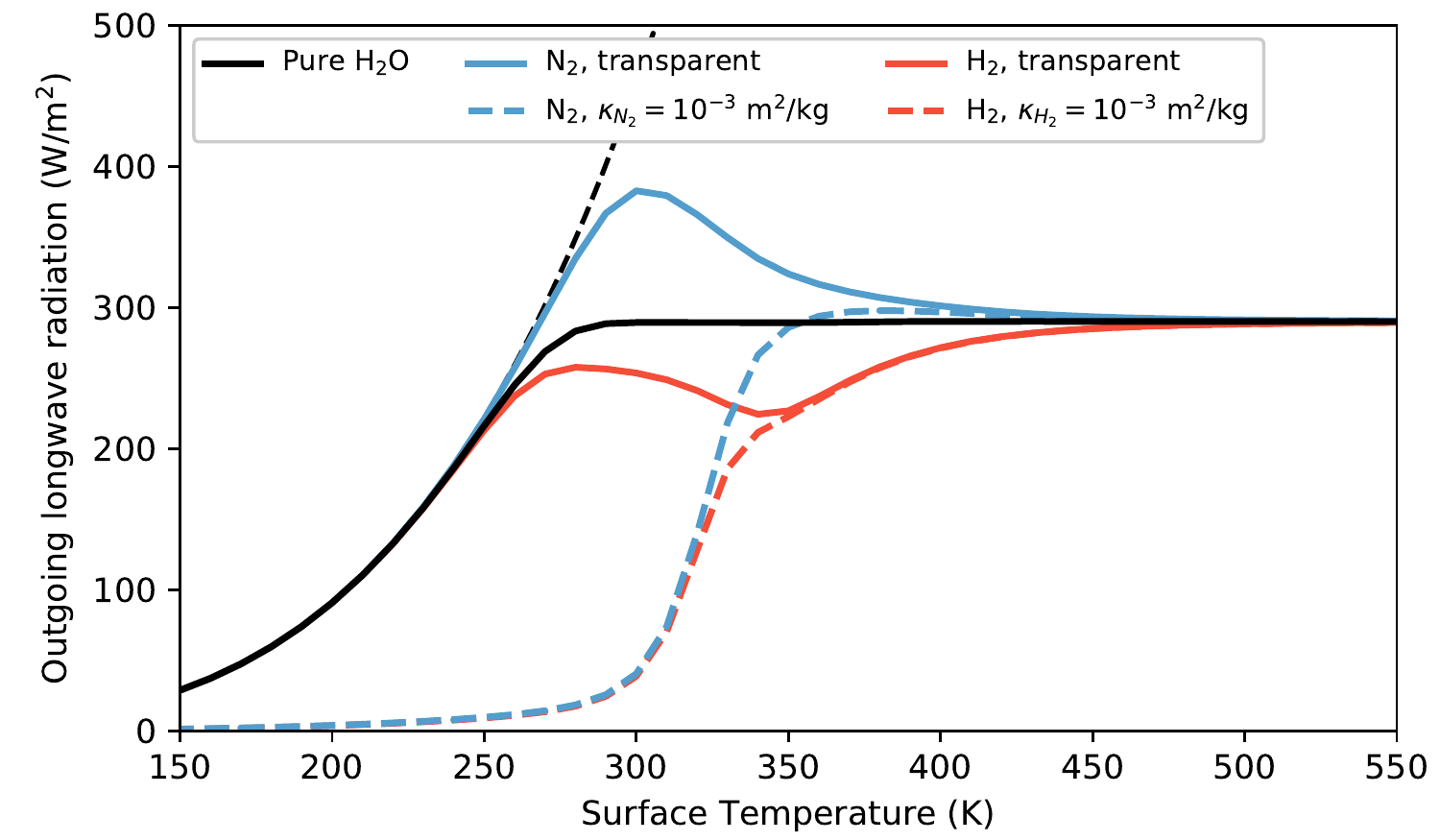}
\caption{The dilute runaway is no longer accessible if the
  background is also a strong grey infrared absorber. Solid lines
  assume that H$_2$O is the only absorbing gas (H$_2$ or N$_2$
  are transparent), dashed lines assume that there is additional
  absorption from the background gas.}
\label{fig:grey02}
\end{figure}

The comparison between Figures \ref{fig:spectral} and \ref{fig:grey}
shows that H$_2$ atmospheres undershoot the steam limit as
our grey model predicts. However, unlike our grey calculations, our
spectrally-resolved calculations do not enter a dilute runaway and,
for H$_2$-rich atmospheres, do not show a non-monotonic OLR with
warming.

The difference between our grey and our spectral calculations arises
from the opacity of the background gas. Whereas our grey calculations
assume that any background H$_2$ is transparent in the infrared, H$_2$
actually has a strong CIA greenhouse effect which is dominant at cold
surface temperatures.
In a H$_2$-rich atmosphere the impact of H$_2$O is thus first felt via
the impact of H$_2$O condensation on the lapse rate (Section
\ref{sec:souffle}), which influences the temperature of the emission
level. It is only at relatively high surface temperatures that H$_2$O
starts to be the dominant opacity source at the emission level so that
OLR approaches the steam runaway.

To illustrate how H$_2$'s greenhouse effect affects our grey
calculations, we repeat them with a dry background gas that is opaque in the
longwave.
This is a general proxy for cases in which the
  atmosphere's background greenhouse effect is non-zero but remains
  fixed with surface temperature (e.g., atmospheres
  dominated by H$_2$ or CO$_2$, but also Earth-like atmospheres with a
  mixed N$_2$-CO$_2$ background).
We use the following opacities, $\kappa_{v} = 10^{-2}$
m$^2$ kg$^{-1}$ and $\kappa_{dry} = 10^{-3}$ m$^2$ kg$^{-1}$. With 1
bar of dry background gas, this means the atmosphere has an optical
thickness of about $10$ even without any water vapor.

Figure \ref{fig:grey02} shows why a background
greenhouse gas is sufficient to obscure the dilute runaway and any
non-monotonic OLR evolution with warming. At cold temperatures the atmosphere is
already optically thick, and OLR is substantially lower than the
surface emission $\sigma T_s^4$. Instead of entering the dilute
runaway at around $270$ K, OLR stays low until shooting up at around $300$ K.
The rapid increase occurs because the reduction of the lapse rate with surface
warming (see Section \ref{sec:souffle}) causes the atmosphere's
emission temperature to increase faster than the surface temperature.
With sufficient warming H$_2$O dominates the opacity near the emission
level, at which point calculations with an opaque background become
identical to calculations with a transparent background
(Fig.~\ref{fig:grey02}). Further warming then results in the familiar
asymptote of OLR towards the steam limit.

Figure \ref{fig:grey02} also shows how additional greenhouse gases
affect the initiation of the runaway greenhouse in a high MMW
atmosphere, such as adding CO$_2$ to a N$_2$-rich atmosphere.
With a transparent N$_2$-rich background the OLR overshoot is large and the
OLR ``bump'' occurs at a low temperature of about 300 K, whereas with
an optically thick N$_2$-rich background the amplitude of the overshoot shrinks
and the OLR bump moves to about 370 K.
Additional greenhouse gases such as CO$_2$ can therefore shift
the initiation of the runaway to higher temperatures in a N$_2$-rich
atmosphere, but they do not alter our result that high MMW atmospheres
experience an OLR overshoot whereas H$_2$-rich atmospheres experience
an OLR undershoot.

In conclusion, H$_2$'s CIA greenhouse explains why the dilute runaway
is not directly apparent in our spectral calculations for H$_2$-rich
atmospheres. Nevertheless, the dilute runaway is still a
useful theoretical limit because it provides a simple
  way of understanding why different background gases lead to an OLR
  overshoot or undershoot, and thus explains why
N$_2$-rich atmospheres abruptly jump from temperate climates into the steam
runaway, whereas H$_2$-rich atmospheres approach the steam runaway smoothly (Figure \ref{fig:spectral}).
Similarly, although H$_2$'s CIA greenhouse prevents OLR from becoming
non-monotonic, the stabilizing climate feedback induced by the Souffl\'e effect is
the underlying reason why H$_2$ atmospheres can remain stable at
extremely elevated surface temperatures. Finally, Souffl\'e dynamics
also shape the remote appearance of H$_2$ atmospheres, which would be
important for interpreting potential transit observations of
H$_2$-rich exoplanets.

%%%%%%%%%%%%%%%%%%%%%%%%%%%%
\section{Discussion}
\label{sec:discussion}

We have shown that H$_2$ leads to a number of novel thermodynamic
feedbacks that allow terrestrial planets with H$_2$-rich atmospheres near
the inner edge of the habitable zone to sustain hot surface
climates.
Our work relates to previous studies on the origin of life and
  the habitability of H$_2$ climates in multiple ways:

\textit{Previous work on the origin of life.}
Our work has interesting consequences for early Earth.
Several bars of H$_2$ are sufficient to create global-mean
surface temperatures above 60$^\circ$ C, either via a direct
H$_2$-H$_2$ greenhouse or through H$_2$'s amplification of the H$_2$O
greenhouse (Fig.~\ref{fig:spectral}).
Such a hot start for life is compatible with the view that the last common
ancestor of life on Earth was a thermophile \citep{giulio2003}, and has the
additional benefit that hot surface conditions could have dramatically
accelerated prebiotic chemical reactions \citep{stockbridge2010}.
By the same token life would also experience a hot start on habitable evaporated cores,
i.e., mini-Neptunes which lost their hydrogen envelopes and turned
into super-Earths \citep{luger2015}.
Theoretical predictions and exoplanet observations suggest that this formation pathway is
common \citep{lopez2012a,fulton2018}, so many habitable-zone 
planets would have sustained thick hydrogen envelopes early on.
As a mini-Neptune loses its hydrogen and surface conditions drop below
the critical point of H$_2$O  ($\sim$220 bar, 647 K), the planet would be
able to sustain a liquid ocean.
% %
As long as any remnant H$_2$ persists, however, the planet's
surface would remain hot (Fig.~\ref{fig:spectral}).
Life on H$_2$-rich exoplanets would thus arise under hyperthermal and
reduced conditions, but might subsequently struggle to adapt to the
planet's gradual cooling and oxidation as the atmospheric hydrogen was
lost to space \citep[see][]{wordsworth2012,abbot2015a}.

\textit{Previous work on H$_2$ climates.} 
Our results complement previous work on the impact of
H$_2$ in colder climates, where its strong CIA greenhouse effect is
dominant.
Several previous studies considered H$_2$-rich atmospheres but
assumed a fixed dry adiabatic temperature structure in their
calculations \citep{pierrehumbert2011a,wordsworth2012}, which
precluded them from identifying the unusual climate feedbacks and
Souffl\'e dynamics of H$_2$ atmospheres we discuss here.
Other studies included full moist thermodynamics but limited
themselves to atmospheres in which H$_2$ did not dominate the mass of
the assumed background gas
\citep{wordsworth2013b,ramirez2014c,batalha2015,ramirez2017e}, so
these studies could similarly not identify the feedbacks that occur
once H$_2$ becomes the dominant background gas.

Finally, our results could be affected by additional physics as
follows:

\textit{Lifetime of H$_2$ atmospheres.}
Our observation that young H$_2$-rich atmospheres can be sustained
long enough to influence the origin of life is sensitive to a number
of factors.
One important factor is the planet's host star. We considered planets around
Sun-like stars, but planets around M dwarfs would lose
hydrogen much more rapidly due to the elevated XUV emission of young M
dwarfs \citep{penz2008}.
Nevertheless, even in this case some rocky planets
around M dwarfs might outgas sufficient hydrogen during the
solidification of the magma ocean to set the stage for an early
H$_2$-rich climate \citep{elkins-tanton2008}.
Another important factor is the formation time of the planet's
hydrogen envelope. Since H$_2$ is constantly escaping to space, and
since the host star's XUV flux diminishes rapidly over time, any delay
in outgassing would substantially lengthen the lifetime of a H$_2$
atmosphere. For example, this might be possible if hydrogen was
produced by a very late giant impact, several hundred million years
after the star's formation \citep[e.g.,][]{quintana2016}.

\textit{Mixed backgrounds.} 
Although we have focused on pure background compositions, it is straightforward to
extend our results to atmospheres with mixed compositions such as a
primordial H$_2$-He or a H$_2$-N$_2$ atmosphere.
Equation \ref{eqn:tau_ratio} shows that in this case the mass-weighted
mean heat capacity, $\bar{c}_p$, compared to the ratio $L_v/T$ determines whether
the background gas leads to an OLR overshoot or undershoot.
Evaluating this criterion for a wide range of mixed backgrounds, we
find that the OLR undershoot remains robust as long as H$_2$ dominates
the background's overall mass (i.e., as long as H$_2$'s mass mixing ratio exceeds 50-55\%).
In the scenario of early Earth with 1 bar of N$_2$, the addition of slightly more than
1 bar of H$_2$ would thus be sufficient to create an atmosphere with
climate feedbacks that are driven by H$_2$ thermodynamics.

\textit{Stratospheres.}  Our calculations assume an all-adiabat
atmosphere and neglect the radiative impact of a stratosphere.
To test our results we repeated our grey radiative
calculations with a stratosphere. We find that our results are
largely robust, even though the addition of a stratosphere lowers
OLR in the dilute runaway limit while barely affecting OLR in the
steam runaway limit.
The underlying reasons for how a stratosphere affects our grey
calculations are not easily summarized, and will be explained in a forthcoming
paper (Koll \& Cronin, in prep). Here we simply note that there is an
intimate connection between our results for the dilute runaway and
previous work on the Komabayashi-Ingersoll (KI) limit, which is a
runaway limit analogous to the steam runaway but which considers only
the stratosphere's radiative balance \citep{komabayasi1967,ingersoll1969}.

\textit{Additional effects of H$_2$.}  More work is needed to explore
additional effects which we were not able to address in this study.
Previous work on the runaway greenhouse in N$_2$-rich atmospheres
underlined that sub-saturation and clouds can have a large impact on
the runaway greenhouse threshold
\citep{pierrehumbert1995a,leconte2013,yang2013}.  Similarly, moist
convection could develop strong spatial and time variability in
H$_2$-dominated atmospheres \citep{li2015}. These dynamics cannot be
fully represented in 1D equilibrium models. Our study is
therefore only a first look at hot habitable climates in
hydrogen-dominated atmospheres, and should be followed up with 3D modeling.

\acknowledgments

We thank Edwin Kite and Sara Seager for detailed feedback and
helpful discussions, and an anonymous reviewer for valuable comments. This article uses the ApJ
template\footnote{\href{https://github.com/pcubillos/ApJtemplate}{https://github.com/pcubillos/ApJtemplate}}. D.D.B.K.~was supported by a James S.~McDonnell Foundation postdoctoral
fellowship. T.W.C. was supported by NSF grant AGS-1623218.

\bibliography{ZoteroLibrary-Feb2019}

\begin{thebibliography}{52}
\expandafter\ifx\csname natexlab\endcsname\relax\def\natexlab#1{#1}\fi

\bibitem[{Abbot(2015)}]{abbot2015a}
Abbot, D.~S. 2015, The Astrophysical Journal Letters, 815, L3

\bibitem[{Barber {et~al.}(2006)Barber, Tennyson, Harris, \&
  Tolchenov}]{barber2006}
Barber, R.~J., Tennyson, J., Harris, G.~J., \& Tolchenov, R.~N. 2006, Monthly
  Notices of the Royal Astronomical Society, 368, 1087

\bibitem[{Barton {et~al.}(2017)Barton, Hill, Yurchenko, Tennyson, Dudaryonok,
  \& Lavrentieva}]{barton2017}
Barton, E.~J., Hill, C., Yurchenko, S.~N., Tennyson, J., Dudaryonok, A.~S., \&
  Lavrentieva, N.~N. 2017, Journal of Quantitative Spectroscopy and Radiative
  Transfer, 187, 453

\bibitem[{Batalha {et~al.}(2015)Batalha, {Domagal-Goldman}, Ramirez, \&
  Kasting}]{batalha2015}
Batalha, N., {Domagal-Goldman}, S.~D., Ramirez, R., \& Kasting, J.~F. 2015,
  Icarus, 258, 337

\bibitem[{Bell {et~al.}(2015)Bell, Boehnke, Harrison, \& Mao}]{bell2015}
Bell, E.~A., Boehnke, P., Harrison, T.~M., \& Mao, W.~L. 2015, Proceedings of
  the National Academy of Sciences of the United States of America, 112, 14518

\bibitem[{Betts {et~al.}(2018)Betts, Puttick, Clark, Williams, Donoghue, \&
  Pisani}]{betts2018}
Betts, H.~C., Puttick, M.~N., Clark, J.~W., Williams, T.~A., Donoghue, P.
  C.~J., \& Pisani, D. 2018, Nature Ecology \& Evolution, 2, 1556

\bibitem[{Dalgarno \& Williams(1962)}]{dalgarno1962}
Dalgarno, A., \& Williams, D.~A. 1962, The Astrophysical Journal, 136, 690

\bibitem[{Ding \& Pierrehumbert(2016)}]{ding2016}
Ding, F., \& Pierrehumbert, R.~T. 2016, The Astrophysical Journal, 822, 24

\bibitem[{{Elkins-Tanton} \& Seager(2008)}]{elkins-tanton2008}
{Elkins-Tanton}, L.~T., \& Seager, S. 2008, The Astrophysical Journal, 685,
  1237

\bibitem[{Fulton \& Petigura(2018)}]{fulton2018}
Fulton, B.~J., \& Petigura, E.~A. 2018, The Astronomical Journal, 156, 264

\bibitem[{Genda {et~al.}(2017)Genda, Iizuka, Sasaki, Ueno, \&
  Ikoma}]{genda2017}
Genda, H., Iizuka, T., Sasaki, T., Ueno, Y., \& Ikoma, M. 2017, Earth and
  Planetary Science Letters, 470, 87

\bibitem[{Giulio(2003)}]{giulio2003}
Giulio, M.~D. 2003, Journal of Theoretical Biology, 221, 425

\bibitem[{Goldblatt(2015)}]{goldblatt2015a}
Goldblatt, C. 2015, Astrobiology, 15, 362

\bibitem[{Goldblatt {et~al.}(2013)Goldblatt, Robinson, Zahnle, \&
  Crisp}]{goldblatt2013}
Goldblatt, C., Robinson, T.~D., Zahnle, K.~J., \& Crisp, D. 2013, Nature
  Geoscience, 6, 661

\bibitem[{Gordon {et~al.}(2017)Gordon, Rothman, Hill, Kochanov, Tan, Bernath,
  Birk, Boudon, Campargue, Chance, Drouin, Flaud, Gamache, Hodges, Jacquemart,
  Perevalov, Perrin, Shine, Smith, Tennyson, Toon, Tran, Tyuterev, Barbe,
  Cs\'asz\'ar, Devi, Furtenbacher, Harrison, Hartmann, Jolly, Johnson, Karman,
  Kleiner, Kyuberis, Loos, Lyulin, Massie, Mikhailenko, {Moazzen-Ahmadi},
  M\"uller, Naumenko, Nikitin, Polyansky, Rey, Rotger, Sharpe, Sung, Starikova,
  Tashkun, Auwera, Wagner, Wilzewski, Wcis\l{}o, Yu, \& Zak}]{gordon2017}
Gordon, I.~E. {et~al.} 2017, Journal of Quantitative Spectroscopy and Radiative
  Transfer, 203, 3

\bibitem[{Hamano {et~al.}(2013)Hamano, Abe, \& Genda}]{hamano2013a}
Hamano, K., Abe, Y., \& Genda, H. 2013, Nature, 497, 607

\bibitem[{Hu {et~al.}(2015)Hu, Seager, \& Yung}]{hu2015}
Hu, R., Seager, S., \& Yung, Y.~L. 2015, The Astrophysical Journal, 807, 8

\bibitem[{Ingersoll(1969)}]{ingersoll1969}
Ingersoll, A.~P. 1969, Journal of the Atmospheric Sciences, 26, 1191

\bibitem[{Kasting {et~al.}(1993)Kasting, Whitmire, \& Reynolds}]{kasting1993}
Kasting, J.~F., Whitmire, D.~P., \& Reynolds, R.~T. 1993, Icarus, 101, 108

\bibitem[{Koll \& Cronin(2018)}]{koll2018d}
Koll, D. D.~B., \& Cronin, T.~W. 2018, Proceedings of the National Academy of
  Sciences, 201809868

\bibitem[{Komabayasi(1967)}]{komabayasi1967}
Komabayasi, M. 1967, Journal of the Meteorological Society of Japan. Ser. II,
  45, 137

\bibitem[{Kopparapu {et~al.}(2013)Kopparapu, Ramirez, Kasting, Eymet, Robinson,
  {Suvrath Mahadevan}, Terrien, {Domagal-Goldman}, Meadows, \&
  Deshpande}]{kopparapu2013b}
Kopparapu, R.~K. {et~al.} 2013, The Astrophysical Journal, 765, 131

\bibitem[{Lazcano \& Miller(1994)}]{lazcano1994}
Lazcano, A., \& Miller, S.~L. 1994, Journal of Molecular Evolution, 39, 546

\bibitem[{Lebrun {et~al.}(2013)Lebrun, Massol, Chassefi\`ere, Davaille, Marcq,
  Sarda, Leblanc, \& Brandeis}]{lebrun2013}
Lebrun, T., Massol, H., Chassefi\`ere, E., Davaille, A., Marcq, E., Sarda, P.,
  Leblanc, F., \& Brandeis, G. 2013, Journal of Geophysical Research: Planets,
  118, 1155

\bibitem[{Leconte {et~al.}(2013)Leconte, Forget, Charnay, Wordsworth, \&
  Pottier}]{leconte2013}
Leconte, J., Forget, F., Charnay, B., Wordsworth, R., \& Pottier, A. 2013,
  Nature, 504, 268

\bibitem[{Li \& Ingersoll(2015)}]{li2015}
Li, C., \& Ingersoll, A.~P. 2015, Nature Geoscience, 8, 398

\bibitem[{Lopez {et~al.}(2012)Lopez, Fortney, \& Miller}]{lopez2012a}
Lopez, E.~D., Fortney, J.~J., \& Miller, N. 2012, The Astrophysical Journal,
  761, 59

\bibitem[{Luger {et~al.}(2015)Luger, Barnes, Lopez, Fortney, Jackson, \&
  Meadows}]{luger2015}
Luger, R., Barnes, R., Lopez, E., Fortney, J., Jackson, B., \& Meadows, V.
  2015, Astrobiology, 15, 57

\bibitem[{Mlawer {et~al.}(2012)Mlawer, Payne, Moncet, Delamere, Alvarado, \&
  Tobin}]{mlawer2012}
Mlawer, E.~J., Payne, V.~H., Moncet, J.-L., Delamere, J.~S., Alvarado, M.~J.,
  \& Tobin, D.~C. 2012, Phil. Trans. R. Soc. A, 370, 2520

\bibitem[{Nakajima {et~al.}(1992)Nakajima, Hayashi, \& Abe}]{nakajima1992}
Nakajima, S., Hayashi, Y.-Y., \& Abe, Y. 1992, Journal of the Atmospheric
  Sciences, 49, 2256

\bibitem[{Owen \& Wu(2016)}]{owen2016a}
Owen, J.~E., \& Wu, Y. 2016, The Astrophysical Journal, 817, 107

\bibitem[{Penz {et~al.}(2008)Penz, Micela, \& Lammer}]{penz2008}
Penz, T., Micela, G., \& Lammer, H. 2008, Astronomy \& Astrophysics, 477, 309

\bibitem[{Pierrehumbert(1995)}]{pierrehumbert1995a}
Pierrehumbert, R. 1995, Journal of the Atmospheric Sciences, 52, 1784

\bibitem[{Pierrehumbert \& Gaidos(2011)}]{pierrehumbert2011a}
Pierrehumbert, R., \& Gaidos, E. 2011, The Astrophysical Journal, 734, 1

\bibitem[{Pierrehumbert(2010)}]{pierrehumbert2010}
Pierrehumbert, R.~T. 2010, Principles of {{Planetary Climate}} (Cambridge, UK:
  {Cambridge University Press})

\bibitem[{Quintana {et~al.}(2016)Quintana, Barclay, Borucki, Rowe, \&
  Chambers}]{quintana2016}
Quintana, E.~V., Barclay, T., Borucki, W.~J., Rowe, J.~F., \& Chambers, J.~E.
  2016, The Astrophysical Journal, 821, 126

\bibitem[{Rafikov(2006)}]{rafikov2006a}
Rafikov, R.~R. 2006, The Astrophysical Journal, 648, 666

\bibitem[{Ramirez \& Kaltenegger(2017)}]{ramirez2017e}
Ramirez, R.~M., \& Kaltenegger, L. 2017, The Astrophysical Journal Letters,
  837, L4

\bibitem[{Ramirez {et~al.}(2014)Ramirez, Kopparapu, Zugger, Robinson, Freedman,
  \& Kasting}]{ramirez2014c}
Ramirez, R.~M., Kopparapu, R., Zugger, M.~E., Robinson, T.~D., Freedman, R., \&
  Kasting, J.~F. 2014, Nature Geoscience, 7, 59

\bibitem[{Ribas {et~al.}(2005)Ribas, Guinan, G\"udel, \& Audard}]{ribas2005}
Ribas, I., Guinan, E.~F., G\"udel, M., \& Audard, M. 2005, The Astrophysical
  Journal, 622, 680

\bibitem[{Schaefer \& Fegley(2010)}]{schaefer2010}
Schaefer, L., \& Fegley, B. 2010, Icarus, 208, 438

\bibitem[{Segura {et~al.}(2003)Segura, Krelove, Kasting, Sommerlatt, Meadows,
  Crisp, Cohen, \& Mlawer}]{segura2003}
Segura, A., Krelove, K., Kasting, J.~F., Sommerlatt, D., Meadows, V., Crisp,
  D., Cohen, M., \& Mlawer, E. 2003, Astrobiology, 3, 689

\bibitem[{Stamnes {et~al.}(1988)Stamnes, Tsay, Wiscombe, \&
  Jayaweera}]{stamnes1988}
Stamnes, K., Tsay, S.-C., Wiscombe, W., \& Jayaweera, K. 1988, Applied Optics,
  27, 2502

\bibitem[{Stevenson(1999)}]{stevenson1999}
Stevenson, D.~J. 1999, Nature, 400, 32

\bibitem[{Stockbridge {et~al.}(2010)Stockbridge, Lewis, Yuan, \&
  Wolfenden}]{stockbridge2010}
Stockbridge, R.~B., Lewis, C.~A., Yuan, Y., \& Wolfenden, R. 2010, Proceedings
  of the National Academy of Sciences, 107, 22102

\bibitem[{Takai {et~al.}(2008)Takai, Nakamura, Toki, Tsunogai, Miyazaki,
  Miyazaki, Hirayama, Nakagawa, Nunoura, \& Horikoshi}]{takai2008}
Takai, K. {et~al.} 2008, Proceedings of the National Academy of Sciences, 105,
  10949

\bibitem[{Voronin {et~al.}(2010)Voronin, Lavrentieva, Mishina, Chesnokova,
  Barber, \& Tennyson}]{voronin2010}
Voronin, B.~A., Lavrentieva, N.~N., Mishina, T.~P., Chesnokova, T.~Y., Barber,
  M.~J., \& Tennyson, J. 2010, Journal of Quantitative Spectroscopy and
  Radiative Transfer, 111, 2308

\bibitem[{Walker {et~al.}(1981)Walker, Hays, \& Kasting}]{walker1981}
Walker, J. C.~G., Hays, P.~B., \& Kasting, J.~F. 1981, Journal of Geophysical
  Research, 86, 9776

\bibitem[{Wordsworth(2012)}]{wordsworth2012}
Wordsworth, R. 2012, Icarus, 219, 267

\bibitem[{Wordsworth \& Pierrehumbert(2013)}]{wordsworth2013b}
Wordsworth, R., \& Pierrehumbert, R. 2013, Science, 339, 64

\bibitem[{Yang {et~al.}(2013)Yang, Cowan, \& Abbot}]{yang2013}
Yang, J., Cowan, N.~B., \& Abbot, D.~S. 2013, The Astrophysical Journal
  Letters, 771, L45

\bibitem[{Yang {et~al.}(2016)Yang, Leconte, Wolf, Goldblatt, Feldl, Merlis,
  Wang, Koll, Ding, Forget, \& Abbot}]{yang2016}
Yang, J. {et~al.} 2016, The Astrophysical Journal, 826, 222

\end{thebibliography}

\begin{appendices}

\section{Escape Timescale of H$_2$}
\label{sec:AppendixA}

We first investigate the possibility of direct atmospheric boil-off
\citep{owen2016a}. In this scenario the planet's atmosphere is so
hot and extended that its photosphere (i.e., the altitude at which the
atmosphere absorbs the bulk stellar flux) becomes comparable to the
planet's Bondi radius. The atmosphere then undergoes hydrodynamic
escape that is powered by the broadband absorbed stellar flux, which allows the planet to shed large
amounts of gas over extremely short timescales.
To evaluate this possibility we first compute the planet's Bondi radius,
\begin{eqnarray}
  R_b & = & \frac{G M_p}{2 c_s^2},
\end{eqnarray}
where $M_p$ is the planet's mass, $c_s=\sqrt{\tilde{\gamma} k T/m}$ is the speed of sound in
H$_2$, and $\tilde{\gamma}=1.4$ for a diatomic gas. We compare the Bondi radius to the planet's photospheric
radius. We assume the photosphere is located at $0.1$ bar
(representative of H$_2$'s emission level in the infrared; this
pressure will be larger at shorter wavelengths so our choice is
conservative), and estimate the photospheric radius as 
\begin{eqnarray}
  R_{photo} & = & \ln(\frac{p_s}{0.1\mathrm{~bar}}) \times H_p,
\end{eqnarray}
where $p_s$ is the surface pressure and $H_p = k T/(m g)$ is the
atmospheric scale height.
We find that Earth-sized planets are stable against atmospheric
boil-off. For example, even with $T=1600$ K and $p_s=1000$ bar,
representative of a planet in the late magma ocean stage with a thick
H$_2$ envelope, $R_{photo}/R_b \sim 0.3$.

This leaves hydrodynamic escape powered by the star's extreme ultraviolet (XUV)
flux as the main mechanism capable of eroding an early
hydrogen-rich atmosphere \citep{pierrehumbert2011a, wordsworth2012}. The escape flux for
energy-limited hydrodynamic escape is given by
\begin{eqnarray}
  \phi & = & \frac{\epsilon F_{XUV} R_p}{4 G M_p},
\end{eqnarray}
where $\epsilon \sim 0.3$ is an efficiency factor, $M_p$ and $R_p$ are
the planetary mass and radius, and $F_{XUV}$ is the stellar XUV
flux. Our formulation assumes that the main escaping
species is atomic hydrogen, i.e., H$_2$ is photodissociated in the
upper atmosphere.

\begin{figure}
\centering
\includegraphics[width=0.75\linewidth,clip]{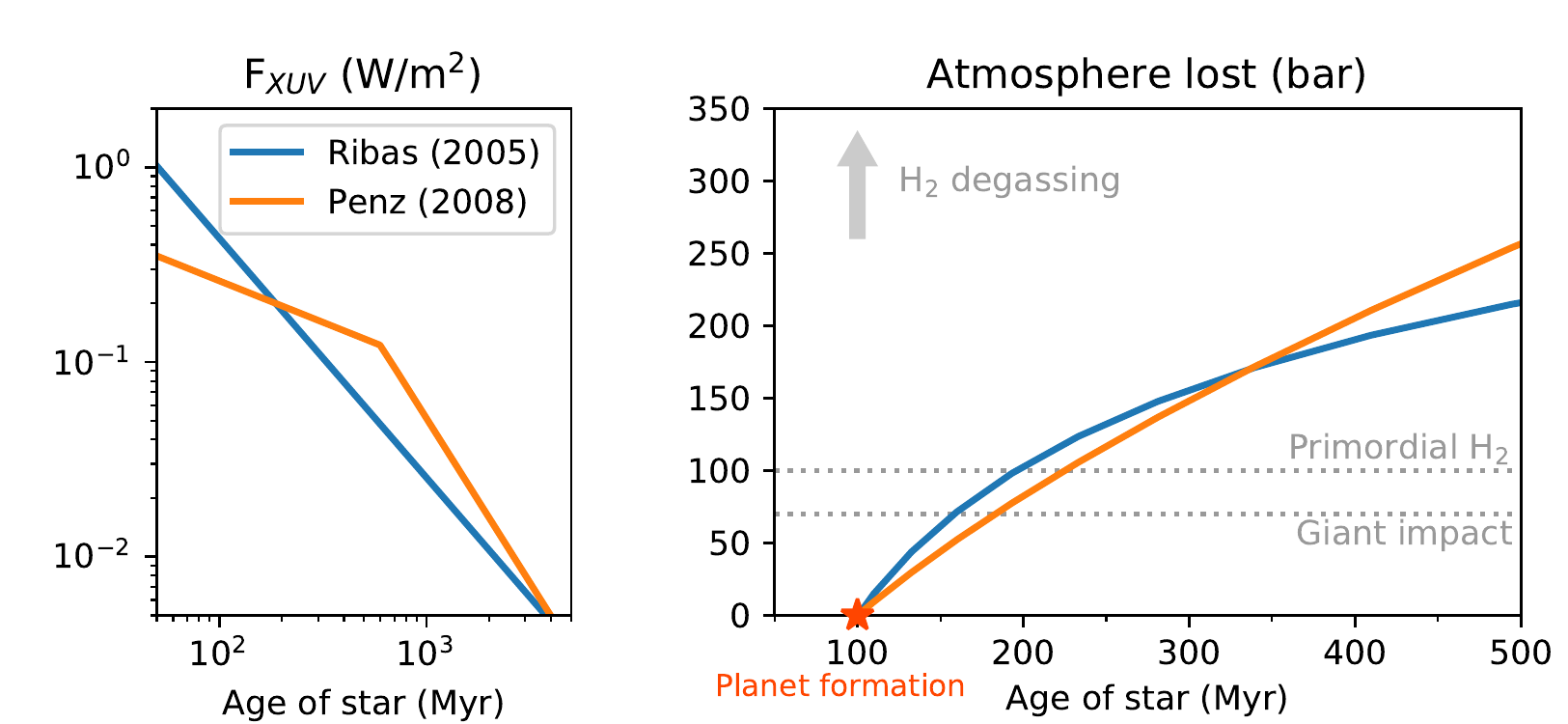}
\caption{Earth-sized planets can retain thick H$_2$ envelopes over 100
  Myr or longer. Left: XUV fluxes of Sun-like
  stars at 1 AU. Right: Atmospheric hydrogen lost for each XUV scaling. Grey
  lines and grey arrow indicate different three mechanisms for
  generating a H$_2$ atmosphere: outgassing from a magma ocean \citep{elkins-tanton2008},
  gravitational capture of primordial H$_2$
  \citep{pierrehumbert2011a}, and oxidation of iron in reaccreting giant-impact
  fragments \citep{genda2017}. }
\label{fig:escape}
\end{figure}

To model the stellar XUV we use the Sun's current XUV flux at
Earth's orbit in the range $\lambda \leq 0.92$ $\mu$m, which is equal
to $F_0=3.9 \times 10^{-3}$ W m$^{-2}$ \citep{ribas2005}. This excludes
Ly$\alpha$ radiation, which does not contribute to the escape flux in the absence of high-altitude
absorbers \citep{pierrehumbert2011a}.
To bracket the time evolution of the stellar XUV flux we use two
different power-laws derived from Solar analogs \citep{ribas2005,penz2008},
\begin{eqnarray}
  F_{XUV} & \propto & t^{-1.23}.
\end{eqnarray}
and
\begin{eqnarray}
  F_{XUV} & \propto 
\begin{cases}
t^{-0.425}, & \mathrm{for~} t\leq 0.6 \mathrm{~Gyr}, \\
t^{-1.69} , & \mathrm{for~} t> 0.6 \mathrm{~Gyr}.
\end{cases}
\end{eqnarray}
We integrate the escape flux, assuming planet formation ends 100 Myr
after the formation of the host star. Figure \ref{fig:escape} shows
how much H$_2$ is lost as a function of time, and how these amounts
compare to different mechanisms for forming an early H$_2$-rich
atmosphere. We find that thick H$_2$-envelopes with 70-100 bars of
H$_2$ can be retained over timescales of 60-120 Myr.

%% ------------------------------------

\section{Validation of Shortwave Radiative Transfer}
\label{sec:AppendixB}

\begin{figure}
\centering
\includegraphics[width=0.5\linewidth,clip]{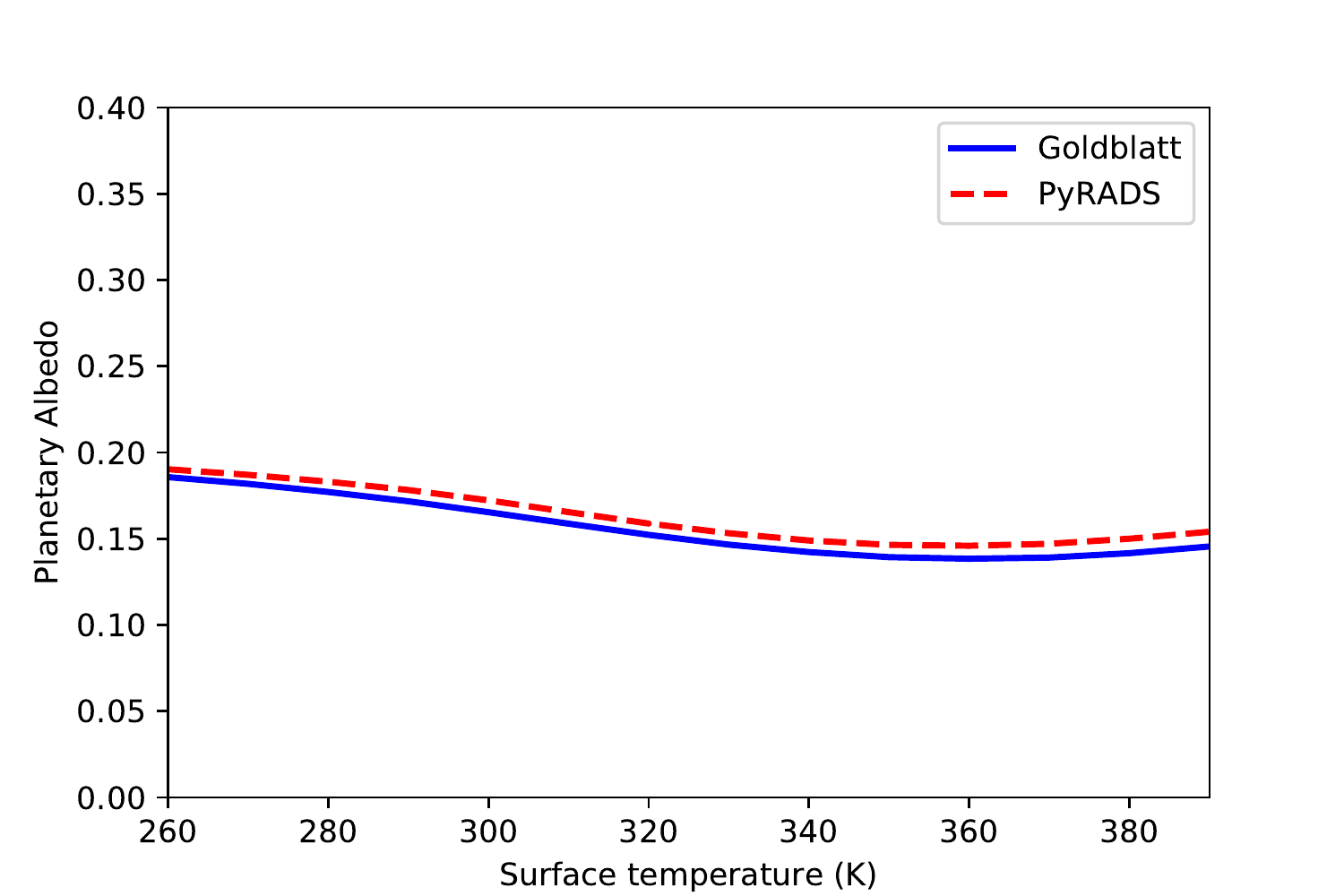}
\caption{Validation of our shortwave calculations. Solid line show
  planetary albedo from a reference set of line-by-line calculations
  \citep{goldblatt2013}, dashed line shows our results. All
  calculations are performed with a surface albedo of 0.12 and 1 bar
  of N$_2$.}
\label{fig:SW_validation}
\end{figure}

Figure \ref{fig:SW_validation} shows that our code compares favorably
against the line-by-line calculations of \citet{goldblatt2013}. To
focus on the validity of our radiative transfer we use the same
atmospheric temperature and humidity profiles as in Goldblatt et al,
and use PyRADS to compute top-of-atmosphere planetary albedos.
For reference, other radiative transfer codes produce typical
differences of about 1-3\% in the top-of-atmosphere planetary albedo
for identical atmospheric profiles \citep{yang2016}.
We find that the albedo computed with PyRADS agrees with the results
from Goldblatt et al to better than 1\%, even though our albedo values
are systematically higher.

The remaining small differences in albedo are likely
due to different modeling assumptions and the use of different opacity
sources. We use the ExoMol BT2 line list,
which has a more complete coverage than the line list used by
Goldblatt et al, so our computed albedos should be lower in the
near-IR and visible. At the same time we only parameterize H$_2$O
absorption in the UV (see Section \ref{sec:methods}), so we could be
underestimating atmospheric absorption at short
wavelengths. In addition, we have found that changes in the assumed
stellar source function can modify our results by up to several
percent, which would be large enough to explain the remaining offset
in Figure \ref{fig:SW_validation}.

\end{appendices}

\end{document}